\documentclass[aps,prd,floatfix,showpacs,10pt,nofootinbib]{revtex4-1}

\usepackage{amsmath,amssymb,amsfonts,amsthm} 
\usepackage{color}
\usepackage{float}
\usepackage{graphicx}
\usepackage{pstricks,pstricks-add}
\usepackage{pst-plot}
\usepackage[scriptsize,nooneline,hang]{caption}
\usepackage[hang,nooneline,scriptsize]{subfigure}

\definecolor{dark-green}{rgb}{0,0.7,0}
\definecolor{dark-blue}{rgb}{0,0.2,0.5}
\definecolor{med-blue}{rgb}{0,0.7,1}
\definecolor{mblue}{rgb}{0,0.2,1}
\definecolor{cnc}{rgb}{0.8,0,0}
\definecolor{light-red}{rgb}{1,0.8,0.8}
\definecolor{dark-yellow}{rgb}{1,0.8,0}
\definecolor{light-blue}{rgb}{0.8,0.9,1}
\definecolor{grey}{rgb}{0.211,0.211,0.211}
\definecolor{verylight-blue}{rgb}{0.93,0.95,1}
\definecolor{light-yellow}{rgb}{1,0.9,0.8}

\newcommand{\weglassen}[1]{}

\begin{document}

\title{Geodesic motion in the space-time of a non-compact boson star}

\author{Valeria Diemer $^{(a)}$ \footnote{n\'ee Kagramanova}}
\email{valeriya.diemer@uni-oldenburg.de}

\author{Keno Eilers $^{(a)}$ }
\email{keno.eilers@uni-oldenburg.de}

\author{Betti Hartmann $^{(b)}$ }
\email{b.hartmann@jacobs-university.de}

\author{Isabell Schaffer $^{(b)}$ }
\email{i.schaffer@jacobs-university.de}

\author{Catalin Toma $^{(b)}$ }
\email{c.toma@alumni.jacobs-university.de}

\affiliation{
$(a)$ Institut f\"ur Physik, Universit\"at Oldenburg, 26111 Oldenburg, Germany\\
$(b)$ School of Engineering and Science, Jacobs University Bremen, 28759 Bremen, Germany}
\date{\today}

\date\today

\begin{abstract}
We study the geodesic motion of test particles in the space-time of non-compact boson stars. These objects
are made of a self-interacting scalar field and -- depending on the scalar field's mass -- can be as dense as 
neutron stars or even black holes. In contrast to the former these objects do not contain a well-defined 
surface, while in contrast to the latter the space-time of boson stars is globally regular, 
can -- however -- only be given numerically. Hence, the geodesic equation also has to be studied numerically. We discuss the
possible orbits for massive and massless test particles and classify them according to the particle's energy and
angular momentum. The space-time of a boson star approaches the Schwarzschild space-time asymptotically, however
deviates strongly from it close to the center of the star. As a consequence, we find 
additional bound orbits of massive test particles close to the center of the star
that are not present in the Schwarzschild case. Our results can be used to make predictions about
extreme-mass-ratio inspirals (EMRIs) and we hence compare our results to recent observational data of the stars
orbiting {\it Sagittarius A$^*$} - the radiosource at the center of our own galaxy. 
\end{abstract}

\maketitle

\section{Introduction}
Solitons play an important r\^{o}le in many areas of physics. As classical solutions of non-linear field theories, they
are localized structures with finite energy, which are globally regular.
In general, one can distinguish between topological and non-topological solitons.
While topological solitons \cite{ms} possess a conserved quantity, the topological charge, that stems (in most
cases) from the spontaneous symmetry breaking of the theory, non-topological solitons \cite{fls,lp} have a conserved Noether
charge that results from a symmetry of the Lagrangian. The standard example of  non-topological solitons
are $Q$-balls \cite{coleman}, which are solutions of theories with self-interacting complex scalar fields. These objects are stationary with an explicitly
time-dependent phase. The conserved Noether charge 
$Q$
is then related to the global phase invariance of the theory and is directly proportional
to the frequency. $Q$ can e.g. be interpreted as particle number \cite{fls}. 
While in standard scalar
field theories, it was shown
that a non-renormalizable $\Phi^6$-potential is necessary \cite{vw}, supersymmetric extensions of the
Standard Model (SM)  also possess $Q$-ball solutions \cite{kusenko}. In the latter case, several scalar fields
interact via complicated potentials. It was shown that cubic interaction terms that result from
Yukawa couplings in the superpotential and supersymmetry (SUSY) breaking terms lead to the existence of $Q$-balls
with non-vanishing baryon or lepton number or electric charge. These supersymmetric
$Q$-balls have been considered as possible candidates for baryonic dark matter 
\cite{dm} and their astrophysical implications have been discussed \cite{implications}.
In \cite{cr}, these objects have been constructed numerically using 
the exact form of a scalar potential that results from gauge-mediated SUSY breaking. However, this
potential is non-differentiable at the SUSY breaking scale.
In \cite{ct} a differentiable approximation of this potential was suggested and the
properties of the corresponding $Q$-balls have been investigated in $3+1$ dimensions.
This was extended to $d+1$ dimensions in \cite{hartmann_riedel2}.  
$Q$-ball solutions with a $\Phi^6$-potential in $3+1$ dimensions have been studied in detail in 
\cite{vw,kk1,kk2}. 

It was realized
that next to non-spinning $Q$-balls, which are spherically symmetric, spinning solutions
exist. These are axially symmetric with energy density of toroidal shape
and angular momentum $J=mQ$, where $Q$ is the Noether charge of the solution
and $m\in \mathbb{Z}$ corresponds to the winding around the $z$-axis. 
Approximated  solutions of the non-linear partial differential equations
were constructed in \cite{vw} by means of a truncated series in the spherical harmonics to describe
the angular part of the solutions. 
The full  partial differential equation was solved numerically in \cite{kk1,kk2,bh}. 
It was also realized in \cite{vw} that in each $m$-sector, parity-even ($P=+1$)
and parity-odd ($P=-1$) solutions exist. Parity-even and parity-odd 
refers to the fact that
the solution is symmetric and anti-symmetric, respectively with respect
to a reflection through the $x$-$y$-plane, i.e. under $\theta\rightarrow \pi-\theta$.
Complex scalar field models coupled to gravity possess so-called ``boson star'' solutions 
\cite{kaup,misch,flp,jetzler,new1,new2,Liddle:1993ha}.
In \cite{kk1,kk2,bh2} boson stars have
been considered that have flat space-time limits in the form of
$Q$-balls. These boson stars are hence self-gravitating $Q$-balls.
In \cite{hartmann_riedel2,hartmann_riedel} the gravitating generalizations of the supersymmetric $Q$-balls
studied in \cite{ct} have been discussed in $d+1$ dimensions. It was found that the behaviour of the
mass and charge at the critical value of the frequency depends crucially on the number of dimensions $d$. 
While in most models considered, the scalar field function is exponentially decaying and hence different
notions of a boson star radius exist, this is different in models with a V-shaped potential \cite{Arodz:2008jk,Arodz:2008nm}. 
In this case, compact boson stars with a well-defined outer radius (very similar to those of ``standard stars'')
can be given \cite{Kleihaus:2009kr,Kleihaus:2010ep,Hartmann:2012da}. 

In this paper we are interested in the possibility to detect boson stars through the motion of massive
and massless test particles in their space-time. In particular, we will be interested in the difference
between test particle motion in a boson star space-time and a Schwarzschild space-time. Since boson stars
are very compact, they have been considered as alternatives to supermassive black holes \cite{schunck_liddle} residing
e.g. in the center of galaxies and geodesic motion of massive test particles 
describing extreme-mass-ratio inspirals (EMRIs) has been discussed in some particular cases in \cite{Kesden:2004qx}.
 While objects with a well-defined surface as alternatives to the
supermassive black hole at the center of our own galaxy, the Milky Way have been ruled out \cite{Broderick:2005xa}, the boson
stars studied in this paper have a scalar field falling of exponentially at infinity and hence strictly speaking
do not have a well-defined outer surface outside which the energy density and pressure, respectively, vanishes. We hence
study {\it non-compact} boson stars in this paper and we will make a detailed analysis of the motion of 
massive and massless test particles in the space-time of such a boson star.

Our paper is organised as follows: in Section II, we give the field theoretical model, the Ansatz and the equations
of motion to describe the space-time of a non-spinning, non-compact boson star.
In Section III we discuss the geodesic equation and give our numerical results in
Section IV. We conclude in Section V.

\section{The space-time of a non-compact boson star}
In the following we will discuss the field theoretical model to describe the 
space-time of a non-spinning, non-compact boson star in which the
test particles will move on geodesics. 

The action $S$ of the field theoretical model reads:
\begin{equation}
 S=\int \sqrt{-g} d^4 x \left( \frac{R}{16\pi G} + {\cal L}_{m}\right)
\end{equation}
where $R$ is the Ricci scalar, $G$ denotes Newton's constant and
the matter Lagrangian is given by
\begin{equation}
\label{lag}
 {\cal L}_{m}=-\partial_{\mu} \Phi \partial^{\mu} \Phi^*
 - U(\vert\Phi\vert)
\end{equation}
where $\Phi$ denotes a complex scalar field and we choose as signature of the metric
$(-+++)$. $U(\vert\Phi\vert)$ is the scalar field potential 
that arises in gauge-mediated supersymmetric breaking in the Minimal Supersymmetric extension of the
Standard Model (MSSM) \cite{cr,ct}:
\begin{equation}
\label{potential}
 U(\vert\Phi\vert)=m^2\eta_{\rm susy}^2 \left(1-\exp\left(-\frac{\vert\Phi\vert^2}{\eta_{\rm susy}^2}\right)\right) \ , 
\end{equation}
where $\eta_{\rm susy}$ corresponds to the scale below which supersymmetry is broken and
is roughly $1$ TeV, while $m$ denotes the scalar boson mass. For $\vert \Phi\vert^2 \ll \eta^2_{\rm susy}$ we 
can give a polynomial expansion of this potential which reads
\begin{equation}
 U(\vert\Phi\vert)=m^2 \vert\Phi\vert^2 - \frac{m^2}{2\eta_{\rm susy}^2} \vert\Phi\vert^4 + 
\frac{m^2}{6\eta_{\rm susy}^4} \vert\Phi\vert^6 + {\cal O}(\vert\Phi\vert^8)  \ .
\end{equation}
Neglecting terms of $\vert\Phi\vert^8$ and higher this potential has the form of the scalar potential
frequently used in the study of $Q$-balls and boson stars  \cite{coleman,vw} and reads
\begin{equation}
\label{phi6}
 \tilde{U}(\vert\Phi\vert)=\lambda\vert\Phi\vert^2\left(\gamma \vert\Phi\vert^4 - \alpha \vert \Phi\vert^2 +\beta\right) \ , 
\end{equation}
where $\lambda$, $\gamma$, $\alpha$ and $\beta$ are constants such that $\sqrt{\lambda \beta}$ corresponds to the
scalar boson star mass $m$. In previous studies, the constants have been fixed to particular values
which fulfill the requirements to find non-trivial solutions to the equations of motion.

The matter Lagrangian ${\cal L}_{m}$ (\ref{lag}) is invariant under the global U(1) transformation
\begin{equation}
 \Phi \rightarrow \Phi e^{i\chi} \ \ \  , \ \ \chi\ \ {\rm constant} \ .
\end{equation}
As such the locally conserved Noether
current $j^{\mu}$, $\mu=0,1,2,3$, associated to this symmetry is given by
\begin{equation}
j^{\mu}
 = -i \left(\Phi^* \partial^{\mu} \Phi - \Phi \partial^{\mu} \Phi^*\right) \  \ {\rm with} \ \ \
j^{\mu}_{; \mu}=0  \ .
\end{equation}
The globally conserved Noether charge $Q$ of the system then reads
\begin{equation}
 Q= -\int \sqrt{-g} j^0 d^3 x  \  .
\end{equation}
Finally, the energy-momentum tensor is given by
\begin{eqnarray}
\label{em}
T_{\mu\nu}&=& g_{\mu\nu} {\cal L} - 2\frac{\partial {\cal L}}{\partial g^{\mu\nu}}\nonumber\\
&=& -g_{\mu\nu} \left[\frac{1}{2} g^{\sigma\rho} 
\left(\partial_{\sigma} \Phi^* \partial_{\rho} \Phi +
\partial_{\rho} \Phi^* \partial_{\sigma} \Phi\right) + U(\Phi)\right] +
\partial_{\mu} \Phi^* \partial_{\nu} \Phi + \partial_{\nu}\Phi^* \partial_{\mu} \Phi\ .
\end{eqnarray}

The coupled system of ordinary differential equations is given by the Einstein
equations
\begin{equation}
\label{einstein}
 G_{\mu\nu}=8\pi G T_{\mu\nu}
\end{equation}
with $T_{\mu\nu}$ given by (\ref{em}) and the Klein-Gordon equation
\begin{equation}
\label{KG}
 \left(\square - \frac{\partial U}{\partial \vert\Phi\vert^2} \right)\Phi=0 \ \   \ \ .
\end{equation}

In the following, we want to study non-spinning boson stars. 
For the metric we use the following Ansatz in isotropic coordinates
\begin{equation}
\label{metric}
 ds^2=-f(r) dt^2 + \frac{l(r)}{f(r)}\left[dr^2 + r^2 d\theta^2 + r^2 \sin^2\theta d\varphi^2\right]  \ .
\end{equation}
For the complex scalar field, we use a stationary Ansatz that contains a periodic dependence of the time-coordinate $t$:
\begin{equation}
\label{ansatz1}
\Phi(t,r)=e^{i\omega t} \phi(r) \ ,
\end{equation}
where $\omega$ is a constant and denotes the frequency.

In order to be able to use dimensionless quantities we introduce the following rescalings
\begin{equation}
\label{rescale}
 r\rightarrow \frac{r}{m} \ \ , \ \ \omega \rightarrow m\omega \ \ ,  \
 \  \phi\rightarrow \eta_{\rm susy} \phi \ \ 
\end{equation}
and find that the equations depend only on the dimensionless coupling constant
\begin{equation}
 \kappa=8\pi G\eta_{\rm susy}^2 = 8\pi \frac{\eta_{\rm susy}^2}{M_{\rm pl}^2}  \ ,
\end{equation}
where $M_{\rm pl}$ is the Planck mass. Note 
that $\kappa$ denotes the ratio between
the SUSY breaking scale and the Planck mass which we expect to be small $\kappa\sim 10^{-31}$
if the SUSY breaking scale is on the order of $1$ TeV. In this paper, we will however also study
larger values of $\kappa$ in order to understand the qualitative behaviour of the solutions
in curved space-time. Note that with these rescalings the scalar boson mass $m$ 
becomes equal to unity.
 
Using the rescalings the Einstein equations read
\begin{equation}
\label{einstein1}
f''=-\frac{1}{2}\frac{1}{fl} \left[ 4\kappa fl^2 \left(1-\exp(-\phi^2)\right) - 8\kappa \omega^2 l^2 \phi^2
+ \frac{4}{r} fl f' - 2l f'^2 + f l' f'\right] 
\end{equation}
and
\begin{equation}
\label{einstein2}
 l''=-\frac{1}{2}\frac{1}{fl}\left[8\kappa l^3 \left(1-\exp(-\phi^2)\right) - 8\kappa \omega^2 \frac{l^3}{f}\phi^2 + \frac{6}{r} f l l' - f l'^2 \right]  \ ,
\end{equation}
while the Klein-Gordon equation for the scalar field function is
\begin{equation}
\label{kg}
 \phi''=-\frac{1}{2}\frac{1}{fl}\left[2\omega^2 \frac{l^2}{f}\phi - 2l^2 \phi \exp(-\phi^2) + \frac{4}{r} fl\phi' + fl'\phi'\right] \ .
\end{equation}
The prime here and in the following denotes the derivative with respect to $r$.

The equations (\ref{einstein1})-(\ref{kg}) have to be solved subject to appropriate boundary conditions. 
These are given by the requirement
of regularity at the origin
\begin{equation}
\label{bc0}
 f'\vert_{r=0} = 0 \ \ , \ \ l'\vert_{r=0} = 0 \ \ , \ \ \phi'\vert_{r=0} = 0
\end{equation}
and by the requirement of finite energy, asymptotically flat solutions
\begin{equation}
\label{bcinf}
 f(r=\infty)=1 \ \ , \ \ l(r=\infty)=1 \ \ , \ \ \phi(r=\infty)=0 \ .
\end{equation}
The set of coupled, nonlinear ordinary differential equations (\ref{einstein1})-(\ref{kg}) has been
studied in detail in \cite{hartmann_riedel}, however in Schwarzschild-like coordinates.
We will quote the relevant results in the following
when necessary, but refer the reader to that paper for details on the field theoretical solutions.

The behaviour of the metric function $f$ determines the mass $M$ of the solution. This is given by \cite{kk1,kk2}
\begin{equation}
 M= \frac{1}{2} \lim_{r\rightarrow\infty} r^2 \partial_r f  \ ,
\end{equation}
while the explicit expression for the Noether charge reads
\begin{equation}
 Q=8\pi \omega \int\limits_{0}^{\infty} \frac{\sqrt{l}}{f} r^2 \phi^2 dr \ .
\end{equation}
Note that the mass $M$ is measured in units of $M_{\rm pl}^2/m$. 

At spatial infinity $r\rightarrow \infty$ the scalar field function decays exponentially
\begin{equation}
 \phi(r >> 1)=\frac{1}{r} \exp(-\sqrt{1-\omega^2} r) + ...  \ .
\end{equation}
Hence $\phi(r)\neq 0$ for $r < \infty$. As such the boson star does not possess a well-defined surface and different
notions of the radius of the boson star are possible (see discussion below). While boson stars with well-defined
surface do exist \cite{Hartmann:2012da} (so-called ``compact boson stars''), which hence rather resemble astrophysical
objects, boson stars without proper surface could play an important r\^{o}le as alternatives to supermassive black holes
in the center of galaxies. 

\section{The geodesic equation}

We consider the geodesic equation
\begin{equation}
 \frac{d^2 x^{\mu}}{d\tau^2} + \Gamma^{\mu}_{\rho\sigma} \frac{dx^{\rho}}{d\tau}\frac{dx^{\sigma}}{d\tau}=0  \ ,
\end{equation}
where $\Gamma^{\mu}_{\rho\sigma}$ denotes the Christoffel symbol given by
\begin{equation}
 \Gamma^{\mu}_{\rho\sigma}=\frac{1}{2}g^{\mu\nu}\left(\partial_{\rho} g_{\sigma\nu}+\partial_{\sigma} g_{\rho\nu}-\partial_{\nu} g_{\rho\sigma}\right)
\end{equation}
and $\tau$ is an affine parameter such that for time--like geodesics $d\tau^2=g_{\mu\nu}dx^{\mu} dx^{\nu}$ corresponds to proper time.

The geodesic Lagrangian $\mathcal{L}_{\rm g}$ for a point particle in the space--time (\ref{metric}) reads
\begin{eqnarray}
\label{lagrangian_geo}
\mathcal{L}_{\rm g}=\frac{1}{2}g_{\mu\nu}\frac{dx^{\mu}}{ds}\frac{dx^{\nu}}{ds}=-\frac{1}{2}\delta
=\frac{1}{2}\left[-f\left(\frac{dt}{d\tau}\right)^{2}+\frac{l}{f}\left(\frac{dr}{d\tau}\right)^{2}+\frac{l}{f} r^2
\left(\frac{d\theta}{d\tau}\right)^{2}+ \frac{l}{f} r^2 \sin^2\theta \left(\frac{d\varphi}{d\tau}\right)^{2}\right]
 \ ,  \end{eqnarray}
where $\delta=0$ for massless particles and $\delta=1$ for massive particles, respectively. 

The constants of motion are the energy $E$ and the angular
momentum $L$ of the particle (absolute value and direction). To fix the direction, we choose
$\theta=\pi/2$. Hence the motion will occur in the equatorial plane. The conserved quantities then are
\begin{eqnarray}
E:=f\frac{dt}{d\tau}\ \ , \ \
L:=\frac{l}{f}r^2 \frac{d\varphi}{d\tau}  \ \ .
\end{eqnarray}
Using these constants of motion we can rewrite (\ref{lagrangian_geo}) as follows
\begin{equation}
\label{geo_eq}
 \left(\frac{dr}{d\tau}\right)^2 = \frac{1}{l}\left(E^2 - V_{\rm eff}(r)\right)  \ ,
\end{equation}
where $V_{\rm eff}(r)$ is the effective potential
\begin{equation}
\label{potential}
 V_{\rm eff}(r)=f\left(\delta + \frac{f}{l} \frac{L^2}{r^2}\right) \ .
\end{equation}
We then find
\begin{equation}
\label{eqphi}
 \varphi - \varphi_0 = 
\int^r_{r_0} \frac{dr}{\sqrt{E^2 - V_{\rm eff}(r)}} \frac{L f}{r^2 \sqrt{l}}  \ .
\end{equation}
The solution to this will give $r(\varphi)$ for a test particle. 

The proper time elapsing along the trajectory is given by the integral \cite{new_reference}
\begin{equation}
\label{proptime}
 \tau -\tau_0 = \int^r_{r_0} \left( \frac{1}{l}\left(E^2 - V_{\rm eff}(r)\right) \right)^{-1/2} dr  \ .
\end{equation}
In the next sections the evolution of the proper time along a geodesic is shown by points.

Since $f$ and $l$ are only given numerically, $\tau(r)$ and $r(\varphi)$ can only be determined numerically.

\section{Numerical results}
We have solved the equations of motion (\ref{einstein1})-(\ref{kg}) numerically using a Newton-Raphson
method with an adaptive grid scheme \cite{colsys}. The relative errors of the numerical integration are on the order of
$10^{-10}-10^{-13}$.  The geodesic equation is integrated
numerically in Fortran using the Fortran Subroutines for
Mathematical Applications--IMSL MATH/LIBRARY. The accuracy of the integration as estimated from application of 
the method for the integration of geodesics in the Schwarzschild space-time is on the order of $10^{-5}$.

\subsection{The metric functions}
In order to be able to compare the space-time of a boson star and that of a Schwarzschild black hole, we 
have first investigated the difference between the metric functions $f$ and $l$. 

\begin{figure}[h!]
\begin{center}
 \includegraphics[scale=0.5]{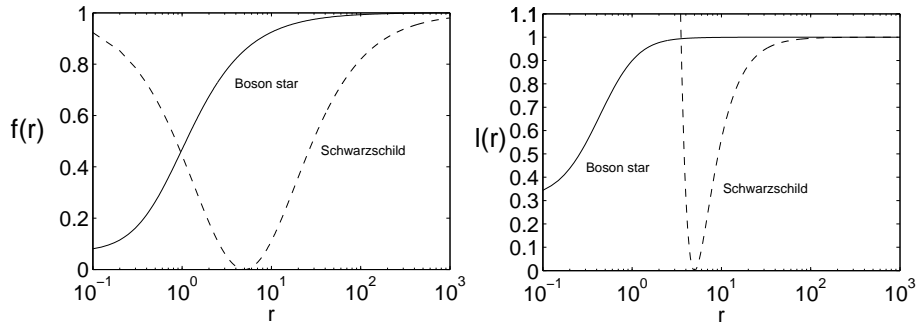}
\end{center}
\caption{We show the metric functions $f(r)$ (left) and $l(r)$ (right) for a boson star with given mass $M\approx 10$
for $\kappa=1.0$.
Also given are the metric functions $f(r)$ and $l(r)$ for a Schwarzschild solution with the same value of $M$.
Note that the Schwarzschild solution has $f(r_h)=l(r_h)=0$ at the event horizon $r_h=M/2$.}
\label{compare_boson_s}
\end{figure}

For that, we have to remind ourselves of how the Schwarzschild metric looks like in isotropic coordinates
which we use in this paper. This is given by (\ref{metric}) with
\begin{equation}
f(r)=\left(\frac{1-\frac{M}{2r}}{1+\frac{M}{2r}}\right)^2 \ \ \ , \ \ \ l(r)=\left(1-\frac{M^2}{4r^2}\right)^2 \ ,
\end{equation}
where $M$ is the ADM mass of the solution and the isotropic coordinate $r$ is related to the Schwarzschild like coordinate $\tilde{r}$ by
$\tilde{r}=r(1+M/(2r))^2$.
In Fig.\ref{compare_boson_s} we show the metric functions $f(r)$ and $l(r)$ for a boson star with given mass $M$ and
in comparison to that the metric functions for a Schwarzschild solution with the same $M$. 

Clearly, the metric functions agree at large $r$, while for small $r$ they deviate strongly. In particular
the metric functions $f(r)$ and $l(r)$ of the Schwarzschild solution tend to zero at the event horizon
$r=r_h=M/2$, while for boson stars $f(r)$ and $l(r)$ stay perfectly regular for all $r\in[0:\infty[$.
This fact will play an important r\^{o}le in the behaviour of test particles at small $r$, while
our results already indicate here that the behaviour at large $r$ will be similar when comparing boson
star space-times with the Schwarzschild space-time. Note that we have chosen a non-realistically high value of
the parameter $\kappa$ here in order to show the qualitative features of the changes. In reality $\kappa$ would
be many orders of magnitude smaller.

\subsection{Mass, charge and radius}
\begin{figure}[h!]
\begin{center}
 \includegraphics[scale=0.5]{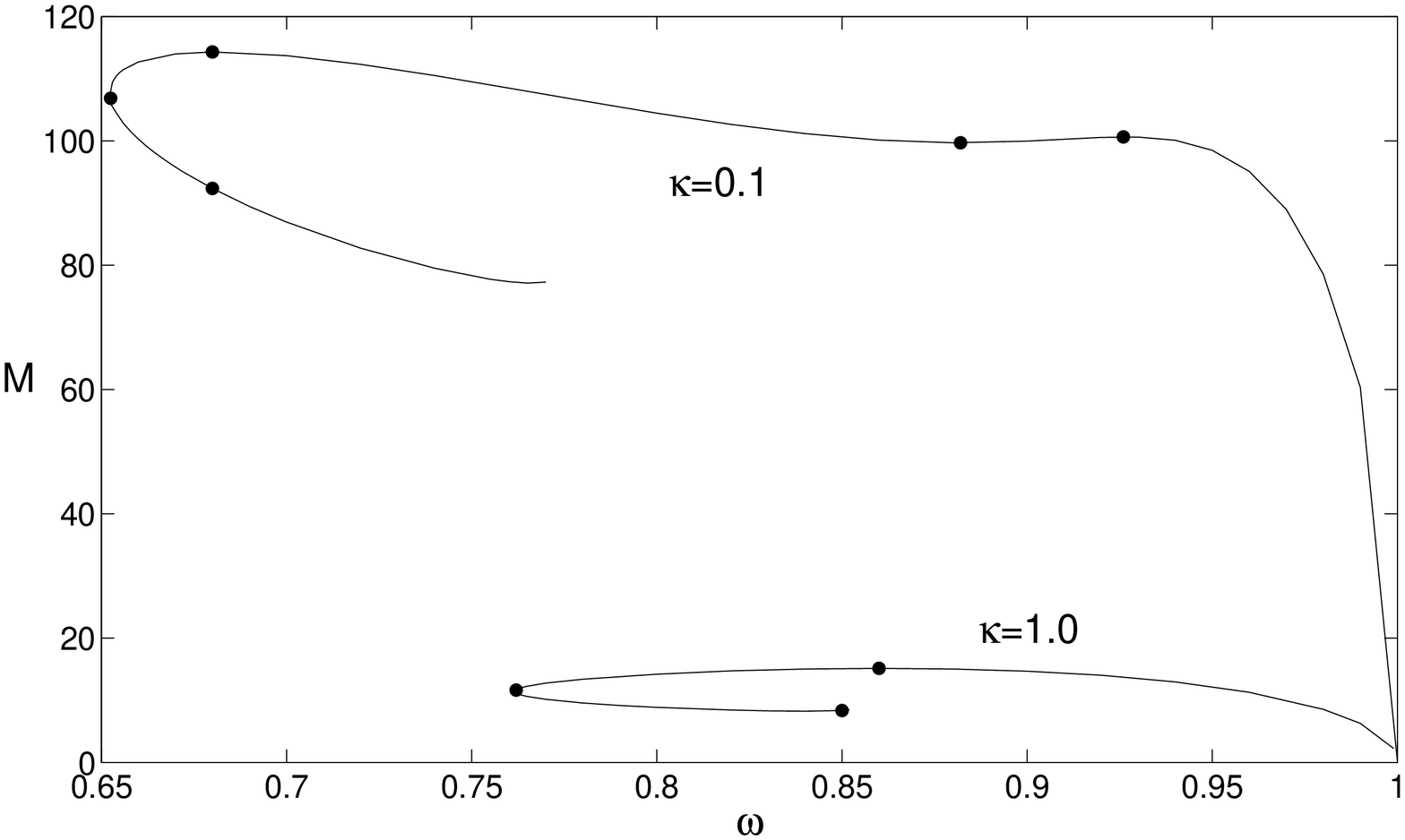}
\end{center}
\caption{The value of the mass $M$ is shown as function of $\omega$ for two different
values of $\kappa$. The black dots on the curves indicate the values of
$\omega$ that we have investigated in more detail and that are given in Table \ref{omega_values}.}
\label{om_m}
\end{figure}

It has been observed before that boson stars exist only in a limited range of the 
frequency $\omega$. At the maximal value of $\omega$ both the mass and the charge of
the boson stars tend to zero, while at the minimal value of $\omega$ a second branch
of solutions starts to form and finally ends at the critical value of $\omega$ after
a number of spirals. In the following, we want to discuss the motion of test particles
in boson star space-times. Here we concentrate on the boson stars at some
particular values of the frequency $\omega$. These are indicated 
in Fig.\ref{om_m}, where we plot the mass $M$ as function of $\omega$ for
$\kappa=0.1$ and $\kappa=1.0$, respectively. The exact numerical values of the mass $M$ and charge $Q$ are given
in Table \ref{omega_values}.

\begin{table}
\begin{center}
  \begin{tabular}{| c | c | c | c | c | c |}
\hline
$\kappa$ & $\omega$ &  $Q$ & $M$ & $\phi(0)$ & $r^{\rm (99)}_{\rm BS}$ \\                
\hline\hline
$0.1$ & $0.9260$ & $101.8523$ & $100.6465$  & $0.4945$ &  $60.4148$     \\ 
$0.1$ & $0.8820$ & $100.8126$ & $99.7029$  & $0.8463$ & $59.9558$     \\
$0.1$ & $0.6800$ (1) & $119.6961$ & $114.3021$ & $2.1891$ & $66.3377$   \\
$0.1$ & $0.6525$ & $108.3567$ & $106.8518$ & $2.7331$ & $63.0380$   \\
$0.1$ & $0.6800$ (2) & $86.4696$ & $92.3413$ & $3.2230$ & $56.3793$   \\
\hline
$1.0$ & $0.8600$ & $15.5600$ & $15.1388$ & $0.2694$ & $80.9035$      \\ 
$1.0$ & $0.7620$ & $11.1205$ & $11.6461$ & $0.7175$ &  $67.1487$   \\
$1.0$ & $0.8500$ & $6.8865$ & $8.3518$  & $1.2718$  &  $52.0577$      \\
\hline
\end{tabular}
\end{center}
  \caption{The values of $\omega$ as indicated by the black dots in Fig.\ref{om_m}
together with the corresponding values of the charge $Q$, mass $M$ and the value $\phi(0)$.
The (1) and (2) denote the solutions on the 1. and 2. branch, respectively. See also \cite{keno_bachelor} for further details
using the potential (\ref{phi6}).}
\label{omega_values}
  \end{table}

In the following, it will also be important to understand what the radius of the boson stars 
and in particular what the extend of the orbits around the boson star are. We would like to discuss
this briefly in the following.
Since the scalar field has an exponential decay in our model, it is difficult
to define a definite radius of the boson stars discussed here. This is
e.g. different in models with a $V$-shaped potential \cite{Hartmann:2012da}, where
a definite radius of the object exists. Hence, different notions of the boson
star radius can be evoked. Either, we can define the radius of the boson
star as the inverse of the scalar boson star mass $r^{\rm (m)}_{\rm BS}=m^{-1}$
(which in our rescaled variables would be equal to unity) or we could define
the radius of the boson star as the radius $r^{\rm (99)}_{\rm BS}$ inside which 
$99\%$ of the mass of the star is contained. It remains to be shown that stars
consisting out of scalar fields can form. Recent results suggest that
a fundamental scalar boson with mass on the order of $m\approx 125$ GeV could exist
in nature \cite{cern_lhc}. If boson stars would consist of this scalar boson, their
typical radii would be (in dimensionful quantities \cite{footnote1}) $\hat{r}^{\rm (m)}_{\rm BS} \approx
10^{-18}$ m. If the star would consist of pions with masses of roughly $140$ MeV 
\cite{new1}, the
typical radius would be $\hat{r}^{\rm (m)}_{\rm BS}\approx 10^{-15}$ m. One could also think
of much lighter scalar particles, e.g. the axion with roughly $10^{-5}$ eV and the dilaton
with $10^{-10}$ eV which would correspond to much larger radii $\hat{r}^{\rm (m)}_{\rm BS}\approx 10^{-2}$ m and 
$\hat{r}^{\rm (m)}_{\rm BS}\approx 10^3$ m, respectively. When computing the radius 
$r^{\rm (99)}_{\rm BS}$ we find that this is roughly two orders of magnitude larger (see Table \ref{omega_values}
for some values). On the other hand, the masses of these boson stars can be quite high.
The mass $M$ is measured in units of $M_{\rm pl}^2/m$. We find that the maximal value of the mass $\hat{M}_{\rm max}$ is well
approximated by
\begin{equation}
 \hat{M}_{\rm max}\approx \frac{7}{4\pi} \frac{M_{\rm pl}^4}{\eta_{\rm SUSY}^2 m}  \ ,
\end{equation}
which should be compared to the typical mass of a neutron star given by $M_{\rm pl}^3/m_{\rm neutron}^2$ with
$m_{\rm neutron}\approx 1 {\rm GeV}$. 

For $m\approx 10^2$ GeV this would correspond
to masses of $10^{38} {\rm GeV}\sim 10^{11} {\rm kg}$ - and this inside a radius of $\hat{r}^{\rm (m)}_{\rm BS} \approx
10^{-18}$ m (or two orders of magnitude higher if we use $r^{\rm (99)}_{\rm BS}$). This is an extremely dense
object, much denser than a typical neutron star. However, these boson stars are stable towards gravitational
collapse due to Heisenberg's uncertainty principle \cite{ms} and due to their extremely high density have been considered as alternatives to supermassive
black holes \cite{schunck_liddle}. The results for radii and masses in dependence on the scalar particle mass
are summarized in Table \ref{table_particles} for a dimensionless value of the boson star mass of $M={\cal O}(10^2)$. 

\begin{table}
\begin{center}
  \begin{tabular}{| c | c | c | c |}
\hline
Particle & $m$ &  $\hat{M}$ & $\hat{r}^{\rm (m)}_{\rm BS}$ \\            
\hline\hline
Higgs & $125$ GeV & $10^{11}$ kg\ & $10^{-18}$ m \\ 
Pion & $140$ MeV & $10^{14}$ kg & $10^{-15}$ m   \\
Axion & $10^{-5}$ eV & $10^{27}$ kg  & $10^{-2}$ m   \\
Dilaton & $10^{-10}$ eV& $10^{32}$ kg &$10^{3}$ m \\
\hline
\end{tabular}
\end{center}
  \caption{The approximate values of the mass $\hat{M}$ and $\hat{r}^{\rm (m)}_{\rm BS}$ of the boson star in dimensionful units
 for different scalar particle with mass $m$ and dimensionless boson star mass $M={\cal O}(10^2)$.
See also \cite{keno_bachelor} for further details using the potential (\ref{phi6}). }
\label{table_particles}
  \end{table}

The relation between the mass $M$ and the charge $Q$ as given in Table \ref{omega_values} also
allows us to make conclusions about the stability of the boson stars. Since $Q$ corresponds to the particle
number, i.e. the number of massive scalar bosons and these have mass $m=1$ (in our rescaled variables), an assembly of $Q$
massive scalar particles would have mass $M=mQ\equiv Q$. Hence, as long as the mass and charge of the
boson star fulfill $M < Q$ they are stable with respect to the decay into $Q$ individual scalar bosons. As such
the boson stars on the first branch of solutions are always stable, while they become unstable on parts of the
second branch.

\subsection{The effective potential and possible orbits}
In all cases, we need to require that $E^2 \geq V_{\rm eff}$ in order to have solutions to (\ref{eqphi}). 
At the intersection points $E^2=V_{\rm eff}$ we have $\frac{dr}{d\tau}=0$. These are the turning points of the motion.
Hence, the effective potential (\ref{potential}) can already give information about the type of
orbits possible.  In the following, we want to discuss which type of orbits are possible in the space-time
of a non-compact boson star. We distinguish different type of orbits as follows: (a) a bound orbit is an
orbit that possesses a minimal and a maximal finite radius, (b) an escape orbit is an orbit
that possesses a minimal finite radius, but extends to infinity, (c) 
a terminating escape orbit is an orbit
whose minimal radius is equal to zero and on which the particle comes from infinity 
to then end at the center of the boson star and (d) a terminating orbit is an orbit with finite maximal
radius and minimal radius equal to zero. Note that when we talk about ``terminating'' in this context
we have the range of the variable $r$ in mind meaning that the particle can reach $r=0$. The possible orbits for massive test particles are summarized in Table
\ref{TypesOfOrbits1}.

Note that in the following we will concentrate on orbits appearing at reasonably small $r$. Since the space-time
of the boson star tends asymptotically to the Schwarzschild space-time, we have in addition all possibilities of
orbits available there at large $r$. We will not discuss these in detail, but refer the reader to the literature 
(see e.g. \cite{schwarz_geo,chandra}).

\begin{figure}[h]
\begin{center}

\subfigure[][$\delta=1$ (1.branch)]{\label{pot_massive_1}
\includegraphics[width=7cm]{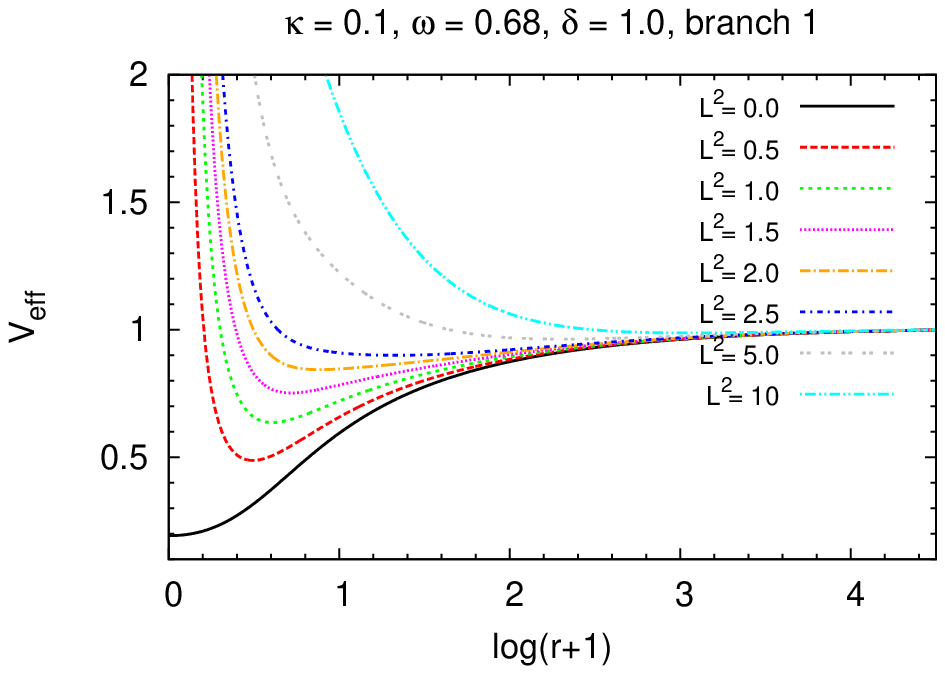}}
\subfigure[][$\delta=0$ (1.branch)]{\label{pot_massless_1}
\includegraphics[width=7cm]{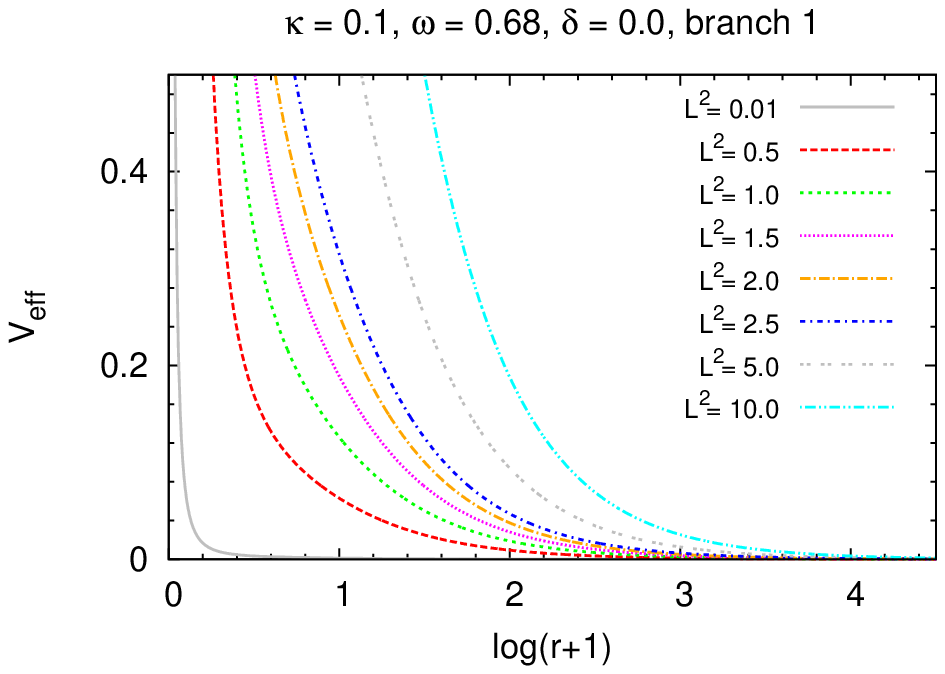}} \\
\subfigure[][$\delta=1$ (2.branch)]{\label{pot_massive_2}
\includegraphics[width=7cm]{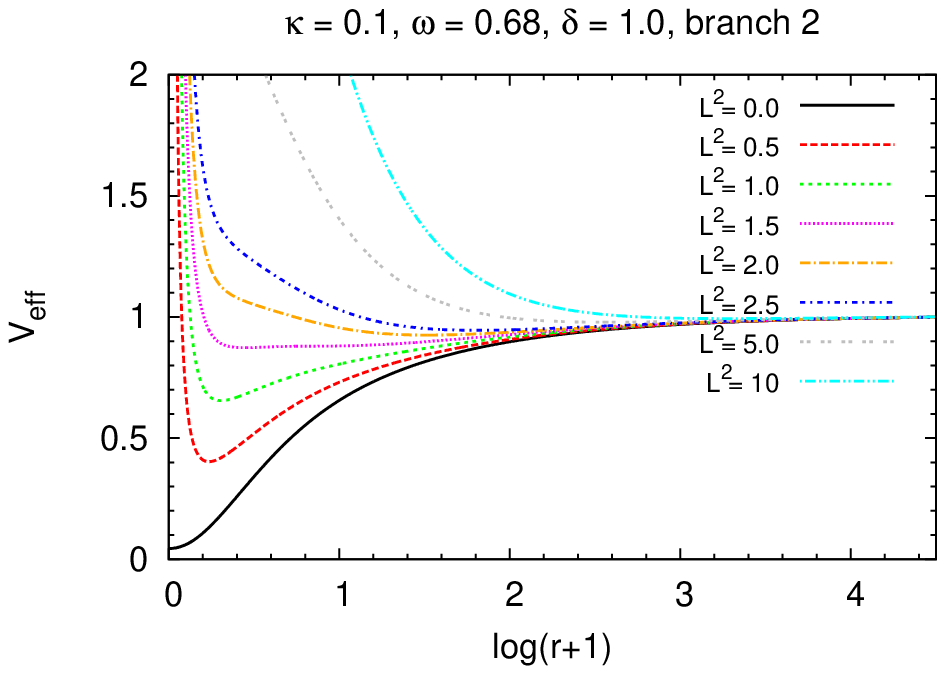}}
\subfigure[][$\delta=0$ (2.branch)]{\label{pot_massless_2}
\includegraphics[width=7cm]{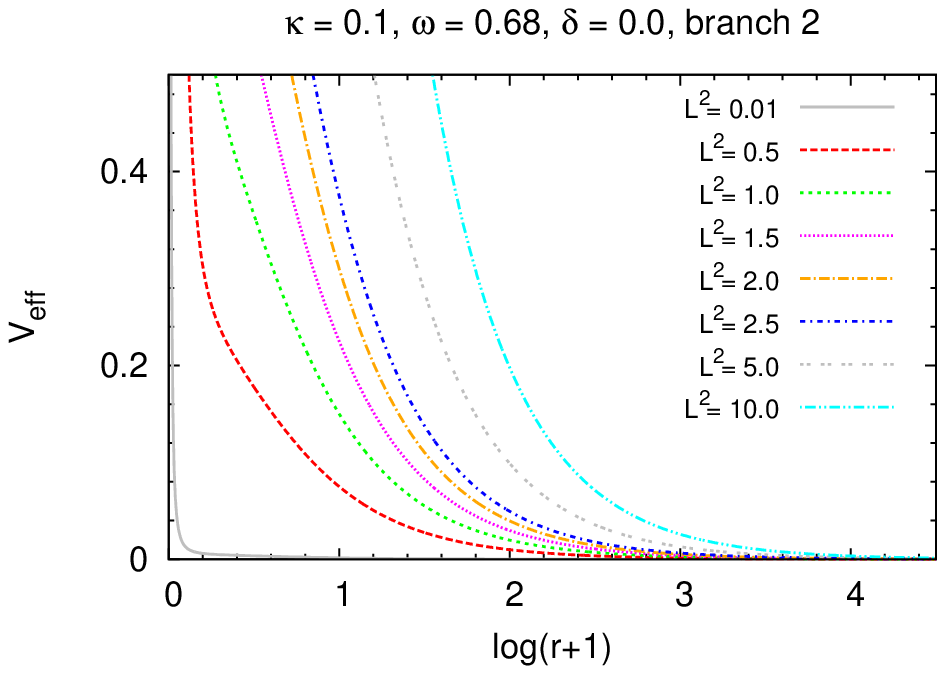}}
\end{center}
\caption{\label{effective} We show the effective potential (\ref{potential}) for massive test particles (left)
and massless test particles (right), respectively, for different values of $L^2$. Here $\kappa=0.1$ and 
$\omega\approx 0.68$. The plots on the left are for the boson star solutions of the 1. branch, the ones on the right
for the boson star solutions of the 2. branch.  }    
\end{figure}

In Fig.\ref{effective} we show the effective potential $V_{\rm eff}$ for massive and massless
test particles, respectively, for a typical boson star space-time with $\kappa=0.1$ and $\omega\approx 0.68$
for the 1. branch as well as for the 2. branch of solutions. 

Again we would like to compare this to the situation in a Schwarzschild space-time. 
First note that $f/l = (1+M/2r)^{-4}$ in the Schwarzschild
case such that for $r\rightarrow r_h=M/2$ the effective potential vanishes since $f$ vanishes. 
As soon as the particle has crossed this horizon we cannot observe it anymore and it will eventually end up in
the physical singularity at $r=0$. For the boson star space-time on the other hand $f$ and $l$ are well behaved
and in particular never become zero in $r\in [0:\infty[$.  As such the angular momentum term $L^2/r^2$ dominates
the effective potential at $r\rightarrow 0$. A particle with $L\neq 0$ can never reach $r=0$ no matter how large
its energy is. Now depending on the massive test particle's energy $E$ and angular momentum $L\neq 0$ it will either
move on a bound orbit or on an escape orbit.  
Clearly, we can have two intersection points between $E^2$ and $V_{\rm eff}$ for massive particles as soon
as the effective potential possesses a local minimum with value smaller than the 
asymptotic value $V_{\rm eff}(r=\infty)=1$ (see Fig.\ref{pot_massive_1} and Fig.\ref{pot_massive_2}).
These two turning points at finite values of $r$ correspond to the minimal radius $r_{\rm min}$
and the maximal radius $r_{\rm max}$ of the bound orbit. For both values of $\kappa$ that we have studied we observe that the further
we move on the branches starting at $\omega=1$, the lower is the value
of the minimum of the effective potential. This can be clearly seen when comparing the effective
potential for a massive test particle with fixed $L^2$ for $\omega=0.68$ on the 1. branch with that
on the 2. branch. This means that
we can have bound orbits for smaller values of $E$ for a given $L$ when moving along
the branches. 
For massive test particles with $L\neq 0$, but $E^2$ large only one intersection point is possible.
These particles will then move on escape orbits. 

On the other hand a radially moving massive particle with $L=0$ can reach $r=0$ if $E^2 \geq V_{\rm eff}(r=0)$
which gives the condition $E^2 - f(0) > 0$. 
Hence, massive test particles without angular momentum
and $E^2 < V_{\rm eff}(r=\infty)=1$ move on  terminating orbits, while those with $E^2 > 1$ move on
terminating escape orbits. See also Table \ref{TypesOfOrbits1} for the possible
orbits of massive test particles.

In Fig.\ref{pot_massless_1} and Fig.\ref{pot_massless_2} we show the effective potentials for 
a massless test particle for different values of $L$ in the space-time of a boson star with $\omega=0.68$. 
We observe that the potential
is monotonically decreasing and that no extrema of the potential exist for $L^2 > 0$, while
$V_{\rm eff}(r)\equiv 0$ for massless test particles without angular momentum, i.e. $L^2=0$.
Hence, we can only have escape orbits for $L^2 > 0$ and terminating escape orbits for $L^2=0$. This 
is summarized in Table \ref{typeoforbits_massless}.  
The difference between the 1. branch solutions and 2. branch solutions does not seem to be very big
for massless test particles.

\begin{table}[h]
\begin{center}
\begin{tabular}{|cccc|}\hline
No. of turning points & $L^2$ 
& range of $r$ & orbit  \\ \hline\hline
 0  &  $L^2=0$ &
\begin{pspicture}(-2,-0.2)(3,0.2)
\psline[linewidth=0.5pt]{->}(-2,0)(3,0)

\psline[linewidth=1.2pt](-2,0)(3,0)
\end{pspicture} 
& terminating escape orbit \\  \hline
 1  & $L^2=0$ &
\begin{pspicture}(-2,-0.2)(3,0.2)
\psline[linewidth=0.5pt]{->}(-2,0)(3,0)
\psline[linewidth=1.2pt]{-*}(-2,0)(0,0)
\end{pspicture} 
 & terminating orbit \\ \hline
 2  & $L^2 > 0$ & 
\begin{pspicture}(-2,-0.2)(3,0.2)
\psline[linewidth=0.5pt]{->}(-2,0)(3,0)
\psline[linewidth=1.2pt]{*-*}(-0.5,0)(1.5,0)
\end{pspicture} 
 & bound orbit \\ \hline
 1 & $L^2 > 0$ &
\begin{pspicture}(-2,-0.2)(3,0.2)
\psline[linewidth=0.5pt]{->}(-2,0)(3,0)
\psline[linewidth=1.2pt]{*-}(0.5,0)(3,0)
\end{pspicture} 
 & escape orbit \\ \hline\hline
\end{tabular}
\caption{Types of orbits possible 
for massive test particles moving in the space-time of a boson star.  
\label{TypesOfOrbits1}}
\end{center}
\end{table}

\begin{table}[h]
\begin{center}
\begin{tabular}{|cccc|}\hline
No. of turning points & $L^2$ 
& range of $r$ & orbit  \\ \hline\hline
 0  &  $L^2=0$ &
\begin{pspicture}(-2,-0.2)(3,0.2)
\psline[linewidth=0.5pt]{->}(-2,0)(3,0)

\psline[linewidth=1.2pt](-2,0)(3,0)
\end{pspicture} 
& terminating escape orbit \\  \hline
 1 & $L^2 > 0$ &
\begin{pspicture}(-2,-0.2)(3,0.2)
\psline[linewidth=0.5pt]{->}(-2,0)(3,0)
\psline[linewidth=1.2pt]{*-}(0.5,0)(3,0)
\end{pspicture} 
 & escape orbit \\ \hline\hline
\end{tabular}
\caption{Types of orbits possible 
for massless test particles moving in the space-time of a boson star.  
\label{typeoforbits_massless}}
\end{center}
\end{table}

For $r \gg 1$ the metric functions of both the boson star space-time and of the Schwarzschild case
approach each other. Hence, the effective potential which depends only on the metric functions
looks the same in both cases and in particular $V_{\rm eff}(r\rightarrow \infty)\rightarrow 1$ for both the
boson star and the Schwarzschild space-time.

\subsection{Examples of orbits}

In the following, we want to show examples of the possible orbits mentioned above.
Bound orbits of massive test particles are of particular interest for two reasons: (a) the perihelion
shift of the orbits can be compared to observational data and (b) black holes are believed to form
accretion discs around their centers and since boson stars have been considered as an alternative
to supermassive black holes something similar should also be possible in that case.
On the other hand, escape orbits of massless test particles, i.e. in particular photons can  be used to compare with observational data of the light deflection.

\subsubsection{Massive test particles}
\begin{figure}[h]
\begin{center}

\subfigure[][bound orbit (1.branch)]{\label{fig_compare1}
\includegraphics[width=7cm]{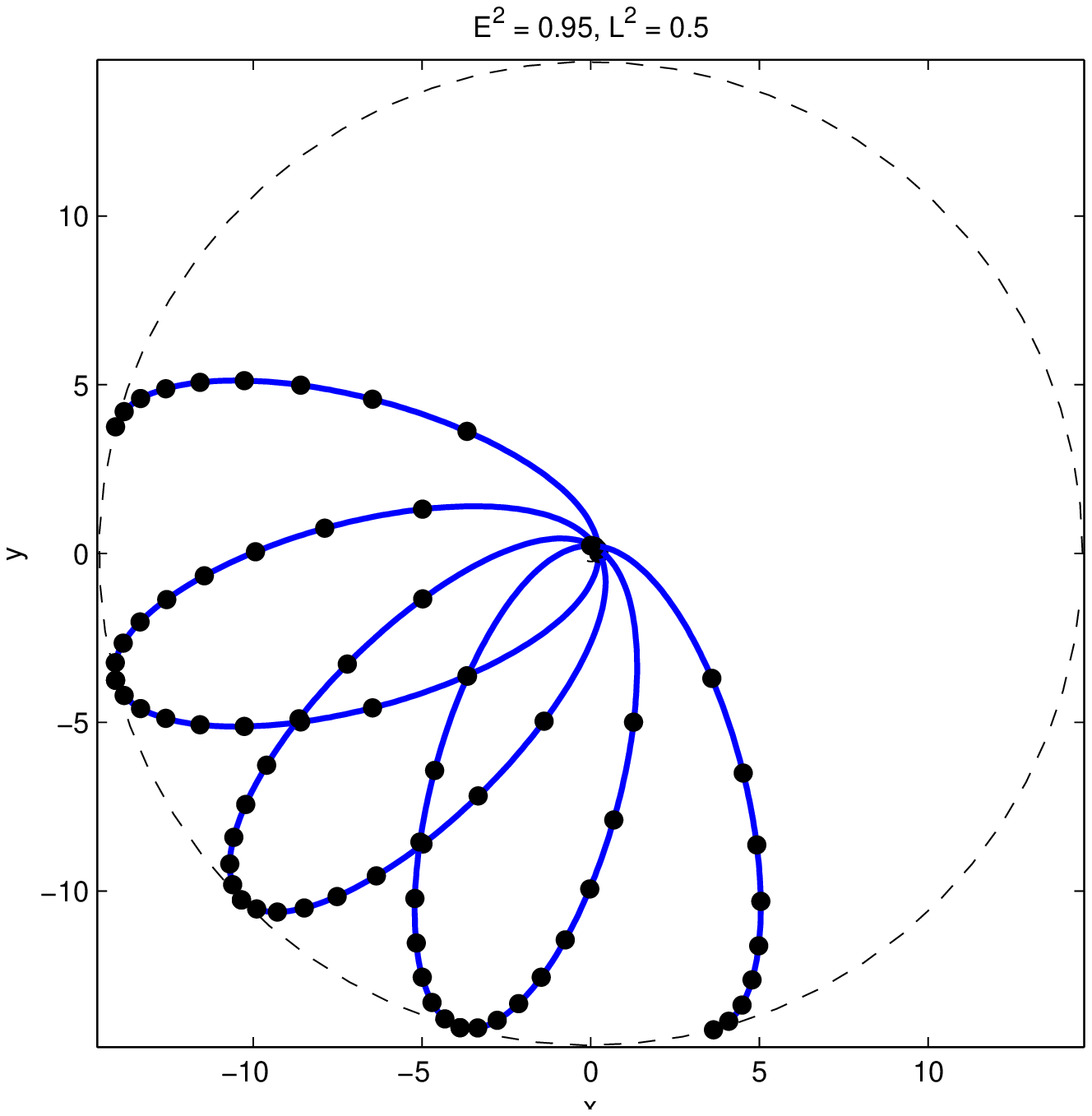}}
\subfigure[][bound orbit (2.branch)]{\label{fig_compare2}
\includegraphics[width=7cm]{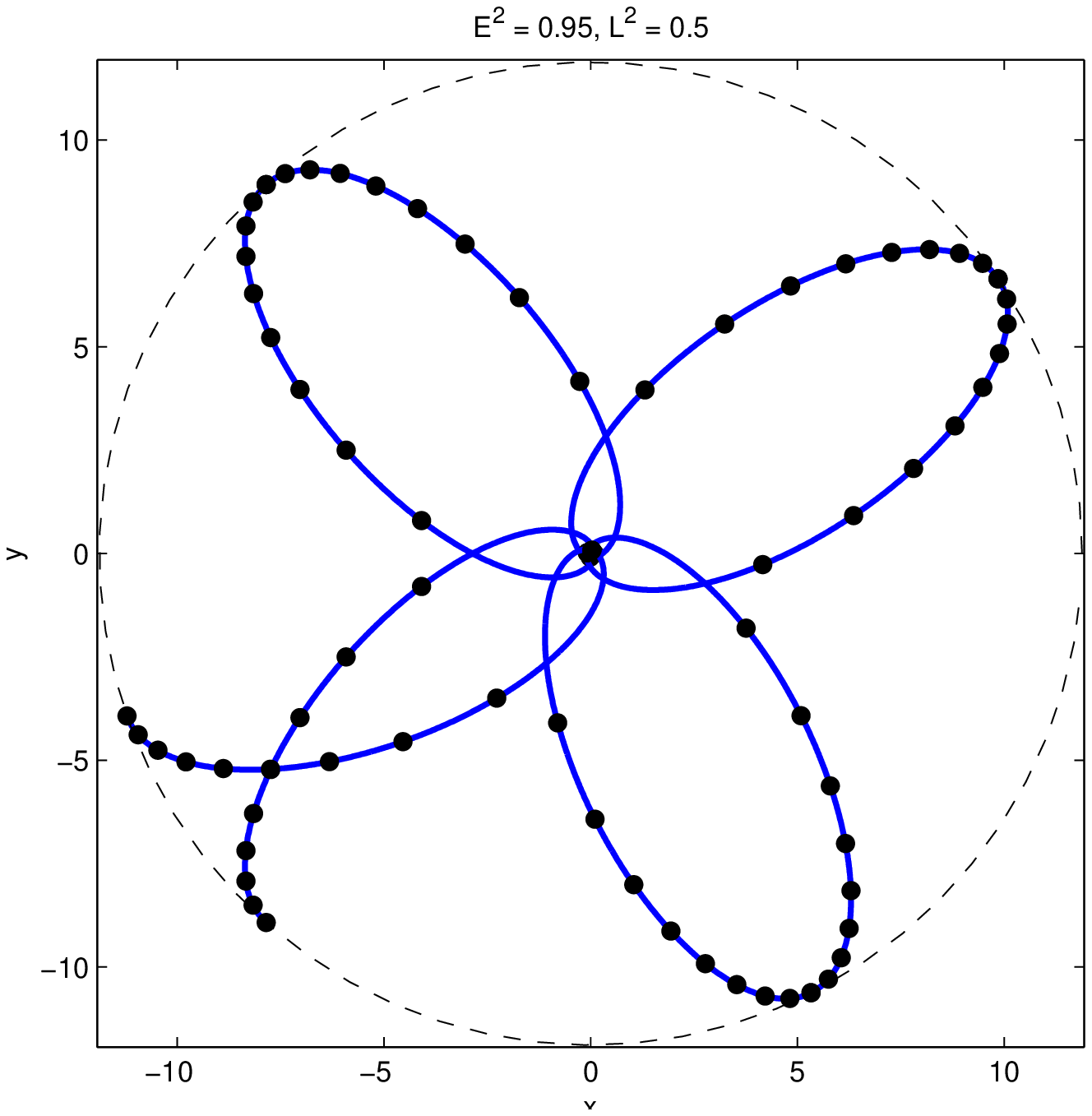}} \\
\subfigure[][escape orbit (1.branch)]{\label{fig_compare5}
\includegraphics[width=7cm]{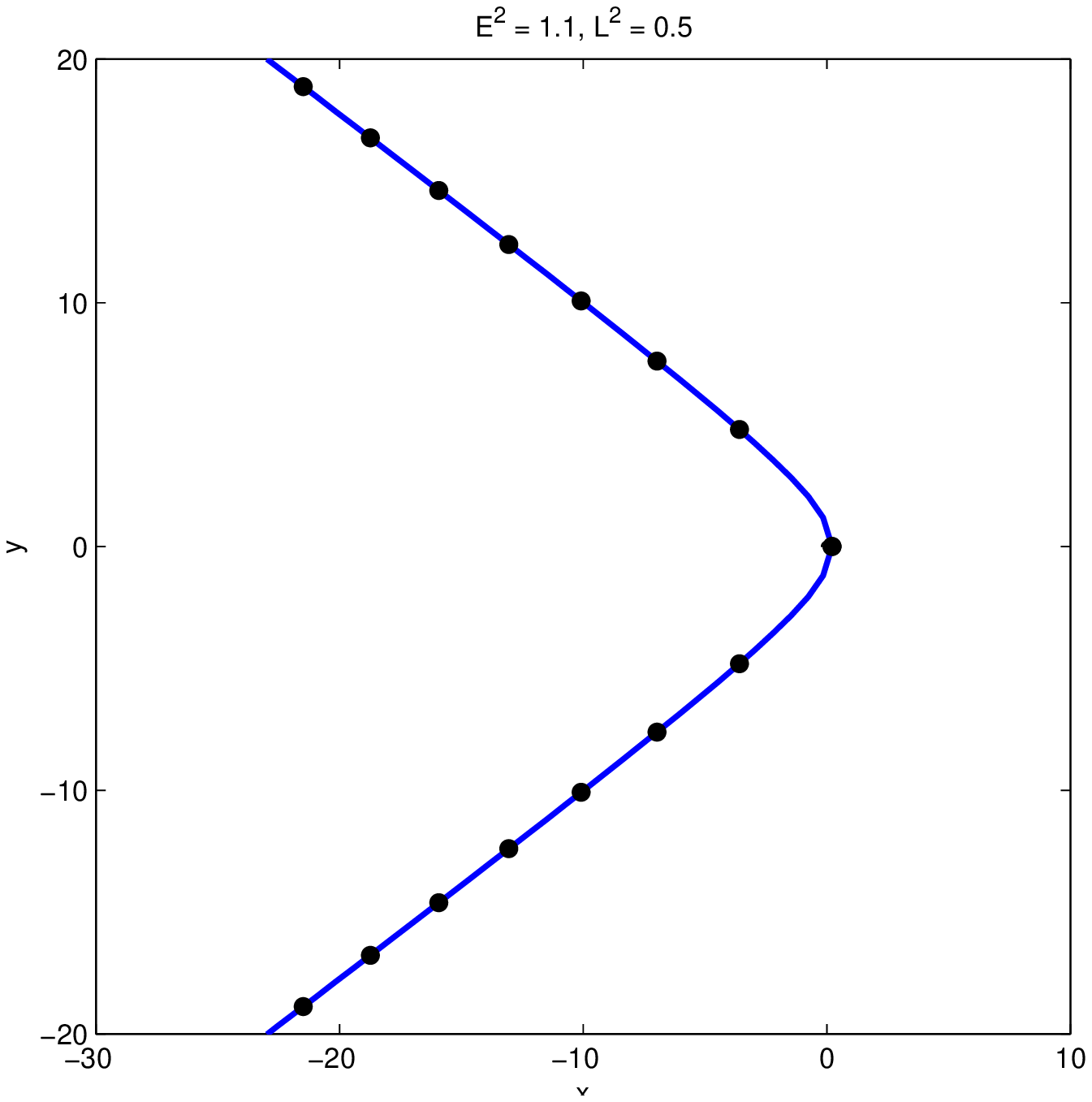}}
\subfigure[][escape orbit (2.branch)]{\label{fig_compare6}
\includegraphics[width=7cm]{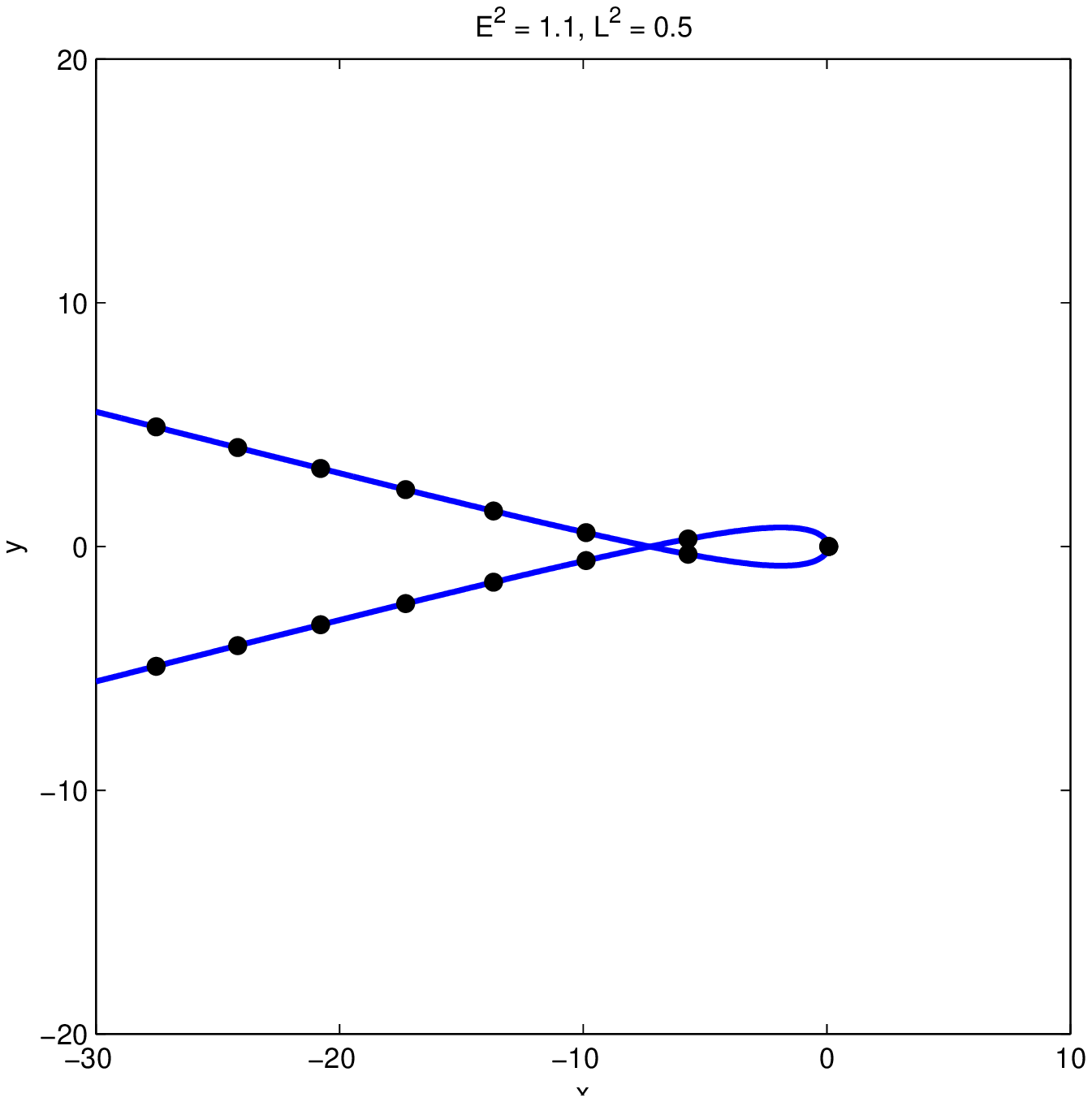}}
\end{center}
\caption{\label{massive_compare_omega} We show bound as well as escape orbits for massive test particles with angular
momentum $L^2=0.5$
in the space-time of a boson star with $\kappa=0.1$ and $\omega=0.68$. For bound orbits the particles
have $E^2=0.95$, while for escape orbits we have chosen $E^2=1.1$. The figures on the left
and right, respectively show the orbits in the space-time of the boson star solution of the 1.branch and 2.branch.
In all cases the center of the boson star is located at the origin of the coordinate system. The evolution of the proper time along the geodesic is shown by points. A test particle covers a larger distance in the same period of proper time, i.e. moves faster, in the vicinity of its perihelion.}    
\end{figure}

In Fig.\ref{massive_compare_omega} we show a bound orbit and an escape orbit of a massive test particle
with $L^2=0.5$ 
in the space-time of a non-compact boson star with $\kappa=0.1$ and
the values of $\omega=0.68$. We compare the motion of massive test particles in the space-time of the boson star
solutions on the 1. and 2. branch, respectively.

For the bound orbits we have chosen $E^2=0.95 < V_{\rm eff}(r=\infty)$, while for the 
escape orbits we have chosen $E^2=1.1 > V_{\rm eff}(r=\infty)$. Though the frequency is the same in both cases,
we see that there are differences in the particle motion. The bound orbits differ clearly in their
perihelion shift. The perihelion shift in the space-time of the boson star of the 2. branch is much larger than
for that of the boson star on the 1. branch (see also discussion below). The maximal radius of the orbit in the case of the 1. branch on the other hand
is slightly bigger (see Table \ref{rmin_rmax} for exact numerical values). 
Moreover, we observe that the further we move on the branches, the lower
is the value of $r_{\rm min}$, while for $\kappa=1.0$ the value of $r_{\rm max}$ 
decreases when moving along the branches. For $\kappa=0.1$ there seems to be a connection
between the value of $r_{\rm max}$ and the value of $M$ with $r_{\rm max}$ being largest
for the maximal possible value of $M$ (see Fig.\ref{om_m}). 

\begin{table}
\begin{center}
  \begin{tabular}{| c | c | c | c | c | c |c|}
\hline
$E^2$ & $L^2$ & $\kappa$ & $\omega$ &  $r_{\rm min}$ & $r_{\rm max}$ & $e$\\                
\hline\hline
$0.97$ & $1.0$ & $0.1$ & $0.9260$ & $1.7688$ &  $19.5923$ & $0.9959$\\
$0.95$ & $1.5$ &$0.1$ & $0.8820$ & $1.8418$   &  $11.3479$ & $0.9867$ \\
$0.95$ & $0.5$ &$0.1$ & $0.6800$ (1) & $0.2376$ & $14.5724$ & $0.9998$\\
$0.95$ & $1.5$ & $0.1$ &  $0.6525$   &  $0.3106$  &  $12.3623$ & $0.9997$  \\
$0.95$ & $0.5$ &$0.1$ & $0.6800$ (2) & $0.0823$ &    $11.8867$ & $0.9999$\\
\hline
$0.97$ & $1.0$ &$1.0$ & $0.8600$ & $0.8302$ &  $27.6181$ & $0.9995$ \\ 
$0.97$ & $1.0$ &$1.0$ & $0.7620$ & $0.2520$ &  $22.3061$ & $0.9999$\\
$0.95$ & $1.0$ &$1.0$ & $0.8500$ & $0.5855$ &  $9.8599$ & $0.9982$\\
\hline
\end{tabular}
\end{center}
  \caption{The values of $r_{\rm min}$, $r_{\rm max}$ as well as of the eccentricity $e:=\sqrt{(r_{\rm max}^2 - r_{\rm min}^2)/r_{\rm max}^2}$ of 
bound orbits with given $E^2$ and $L^2$ in boson
star space-times with $\kappa$ and $\omega$. The $L^2$ value is chosen 
close to the upper boundary up to which bound orbits are possible such that the eccentricity of the
orbits is close to its maximal possible value. }
\label{rmin_rmax}
  \end{table}

The escape orbits also differ in the sense that in the case of the 1. branch the test particles approach the
boson star and get deflected into a direction nearly perpendicular to the original one. 
In the case of the 2. branch the particle encircles the boson star and gets deflected into a direction similar
to the one that it came from.  

\begin{figure}[p!]
\subfigure[$\omega=0.9260$, $L^2=1$, $E^2=0.884$]{\label{iscos1}
\includegraphics[width=6cm]{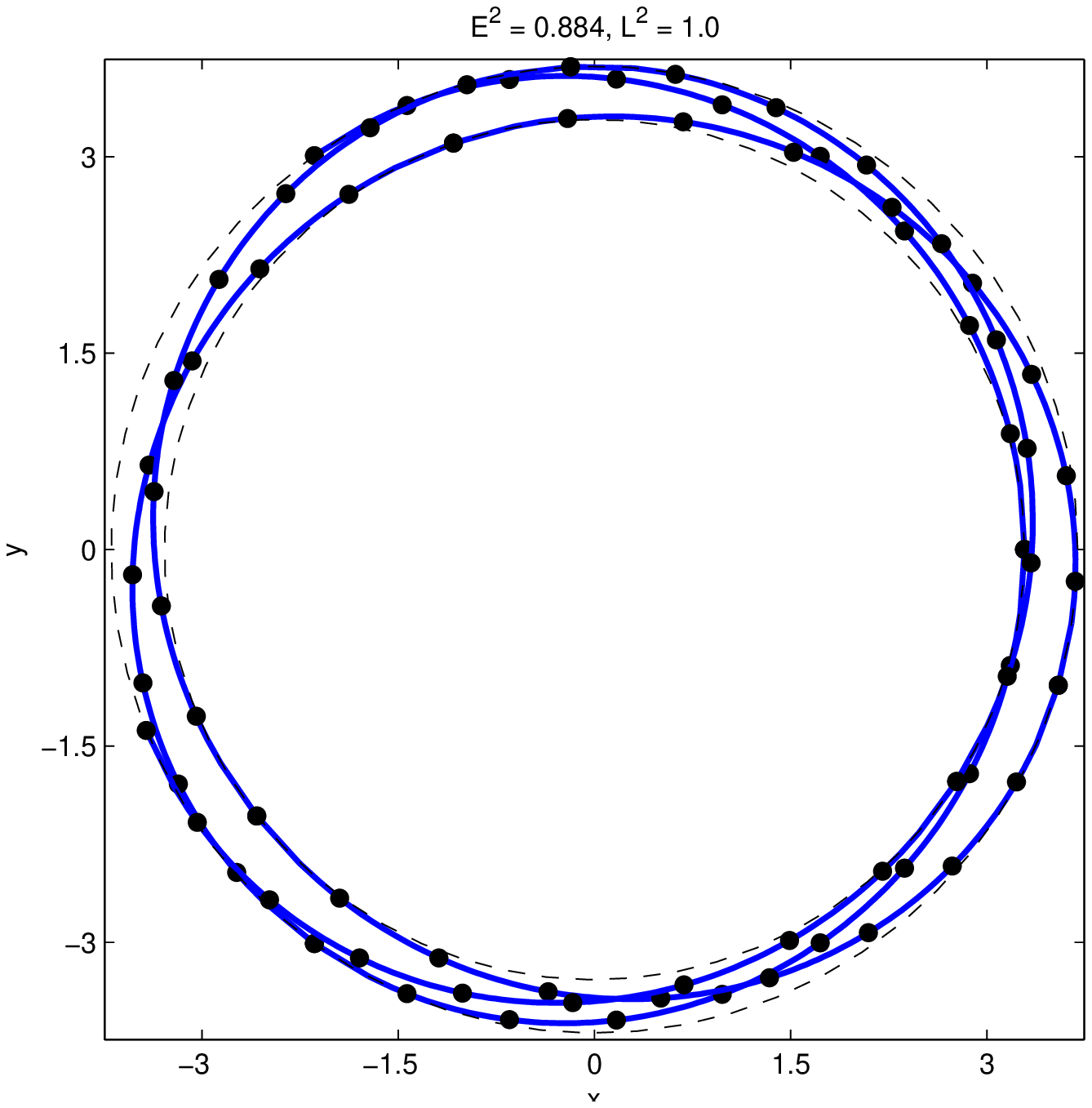}}
\subfigure[$\omega=0.8820$, $L^2=1.5$, $E^2=0.891$]{\label{iscos2}
\includegraphics[width=6cm]{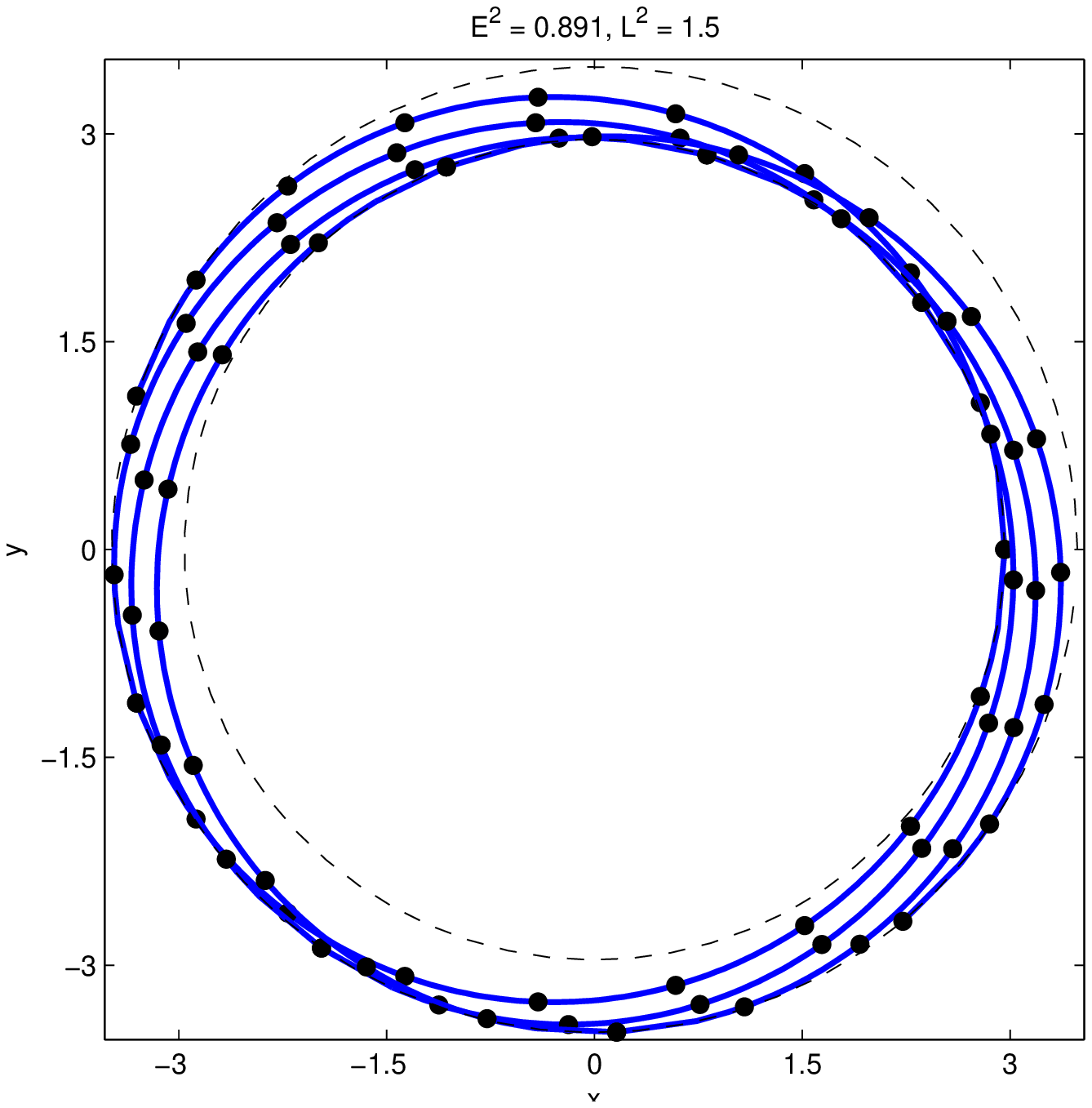}}
\subfigure[$\omega=0.6800$ (1. branch), $L^2=0.5$, $E^2=0.489$]{\label{iscos3}
\includegraphics[width=6cm]{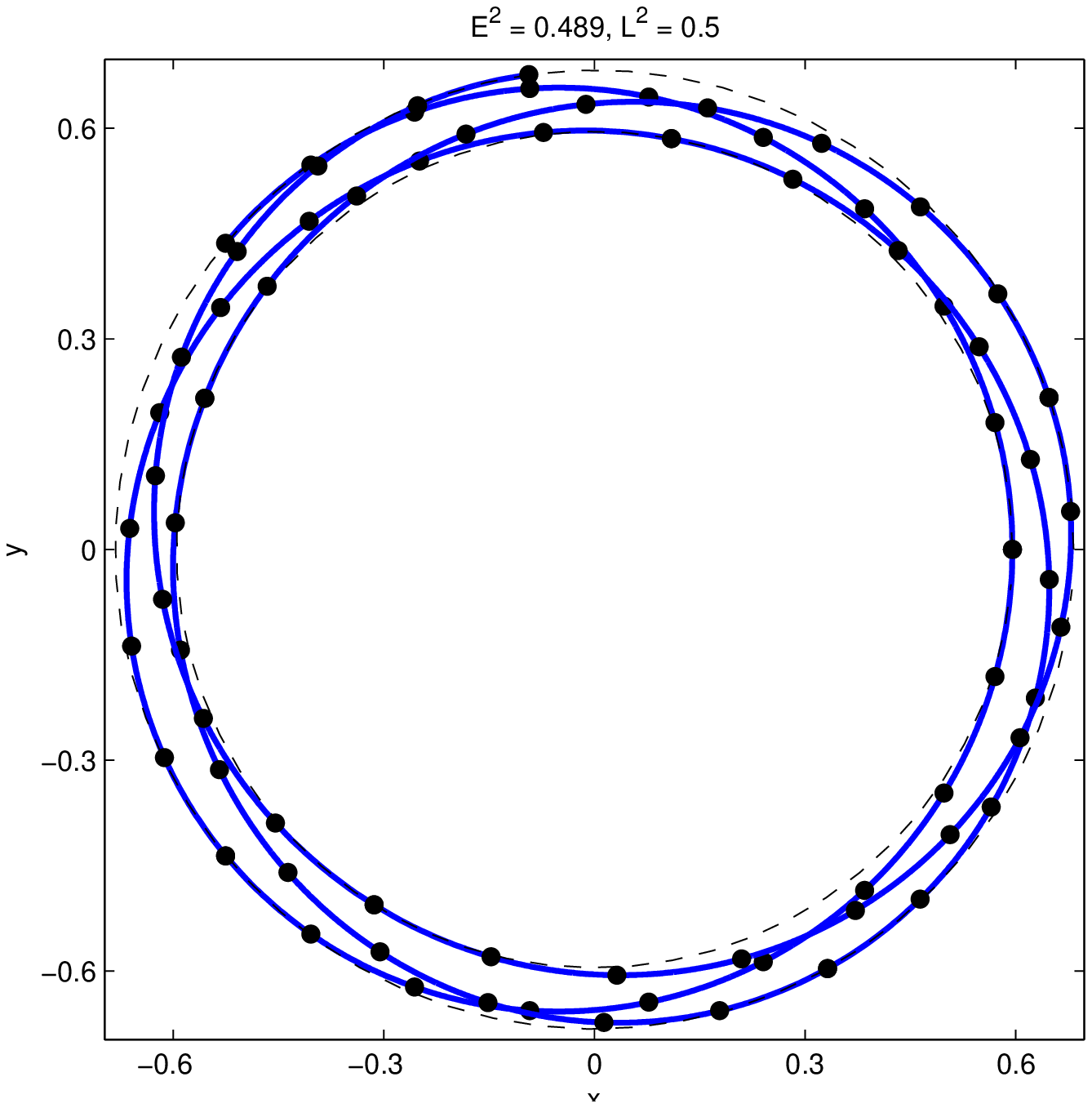}}
\subfigure[$\omega=0.6800$ (2. branch), $L^2=0.5$, $E^2=0.404$]{\label{iscos4}
\includegraphics[width=6cm]{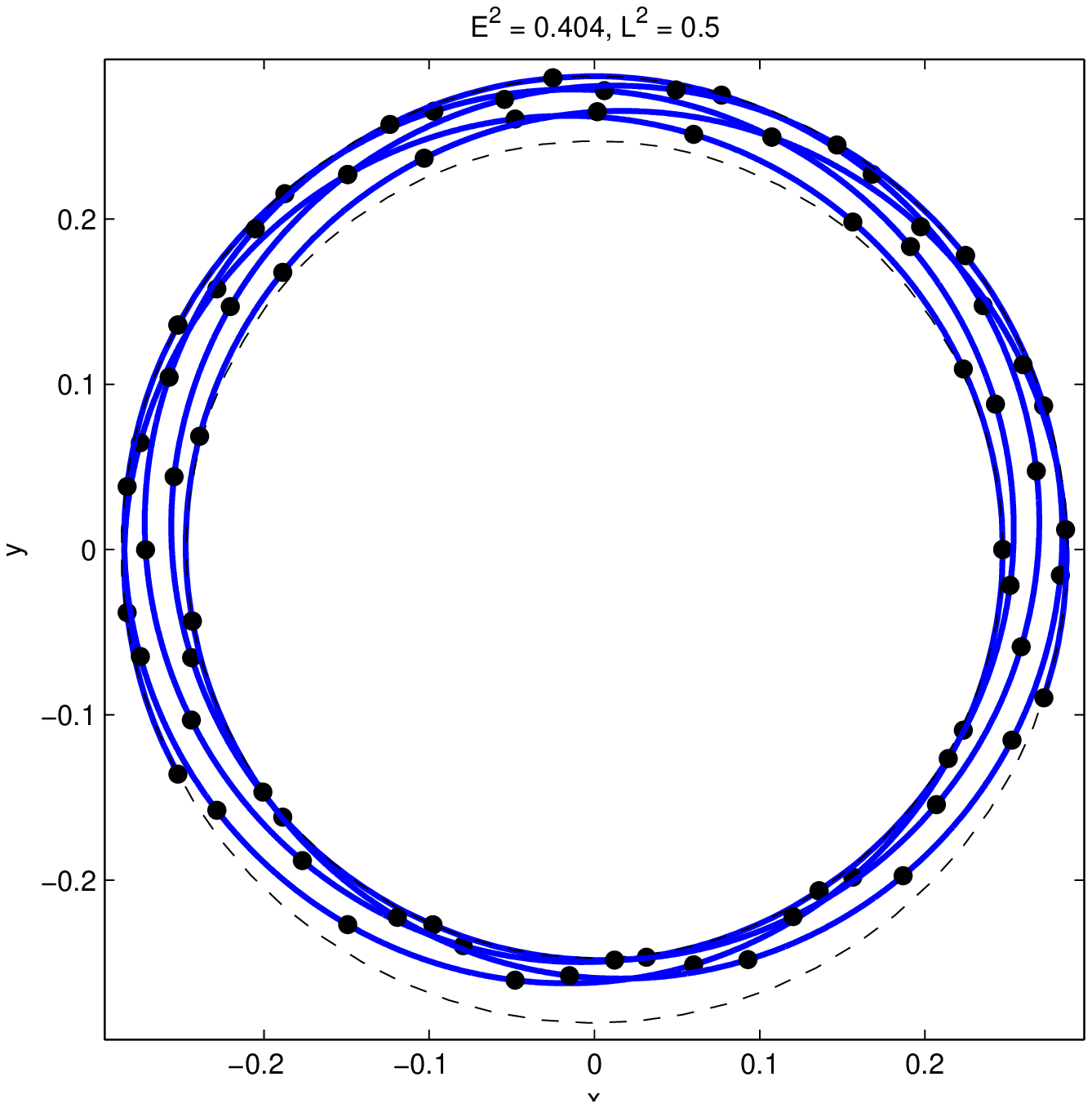}}
\subfigure[$\omega=0.6525$, $L^2=1.5$, $E^2=0.742$]{\label{iscos5}
\includegraphics[width=6cm]{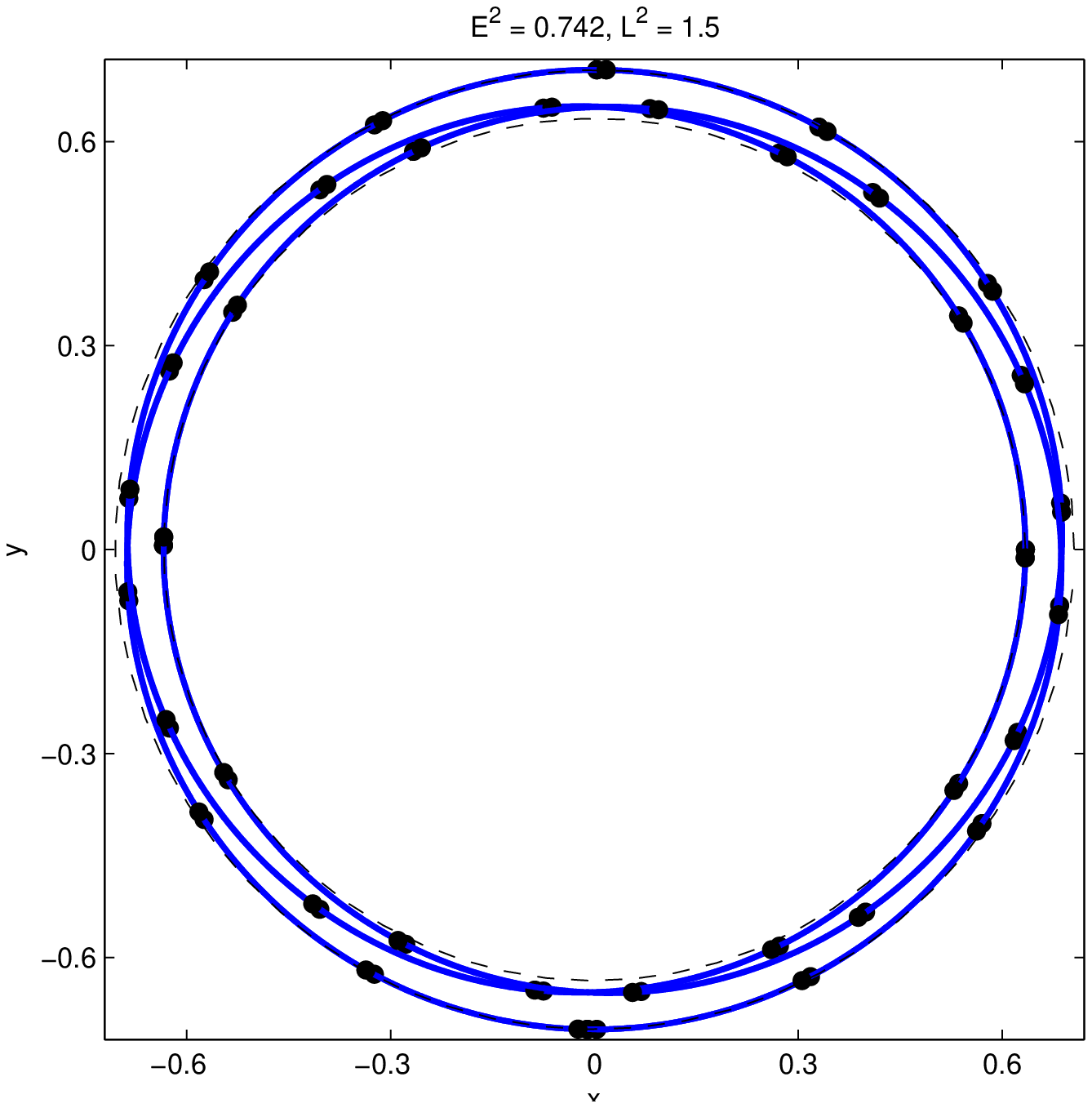}}
\caption{\label{iscos} We show nearly circular, so-called `` stable circular orbits''  (SCOs) 
for massive test particles with given angular momentum $L^2$
in the space-time of a boson star with $\kappa=0.1$ and different values of $\omega$. The evolution of the proper time along the geodesic is shown by points. }    
\end{figure}

\paragraph{Stable circular orbits (SCOs)} 
If the effective potential possesses a local minimum and we choose $E^2$ such that it is equal to
the value of the effective potential at this minimum the orbits are circular orbits which we refer to as
``stable circular orbits'' (SCOs). Examples of these orbits are shown in Fig.\ref{iscos} for different values
of $\omega$ and $L^2$. Note that as soon as we have fixed the $L^2$ the SCO exists only for one specific value 
of $E^2$. 

We observe that the further we move on the branches, the smaller is the radius of the SCO of a particle of given angular
momentum. This can be seen for a test particle with $L^2=1.5$ when comparing Fig.\ref{iscos2} and Fig.\ref{iscos5} as
well as for a particle with $L^2=0.5$ when comparing Fig.\ref{iscos3} and Fig.\ref{iscos4}. Apparently, the extend
of the orbit is not only connected to the mass of the boson star since the SCO of a $L^2=1.5$ particle has
larger radius in the space-time of a boson star with $\omega=0.8820$, $M=99.7029$ as compared to that in a boson
star space-time with $\omega=0.6525$, $M=106.8518$, while the SCO for a $L^2=0.5$ particle has a larger radius
in a boson star space-time with $\omega=0.6800$, $M=114.3021$ as compared to that in a boson star space-time with
$\omega=0.6800$, $M=92.3413$. The message that one should take away here is that the non-trivial scalar field
seems to play an important r\^ole in the particle motion and that no direct relation between the mass of the object
and the SCO exists as one would expect from a Newtonian treatment of the orbits.

\subsubsection{Massless test particles}

\begin{figure}[p!]
\subfigure[$\omega=0.9260$, $\kappa=0.1$]{\label{mo1}
\includegraphics[width=5.7cm]{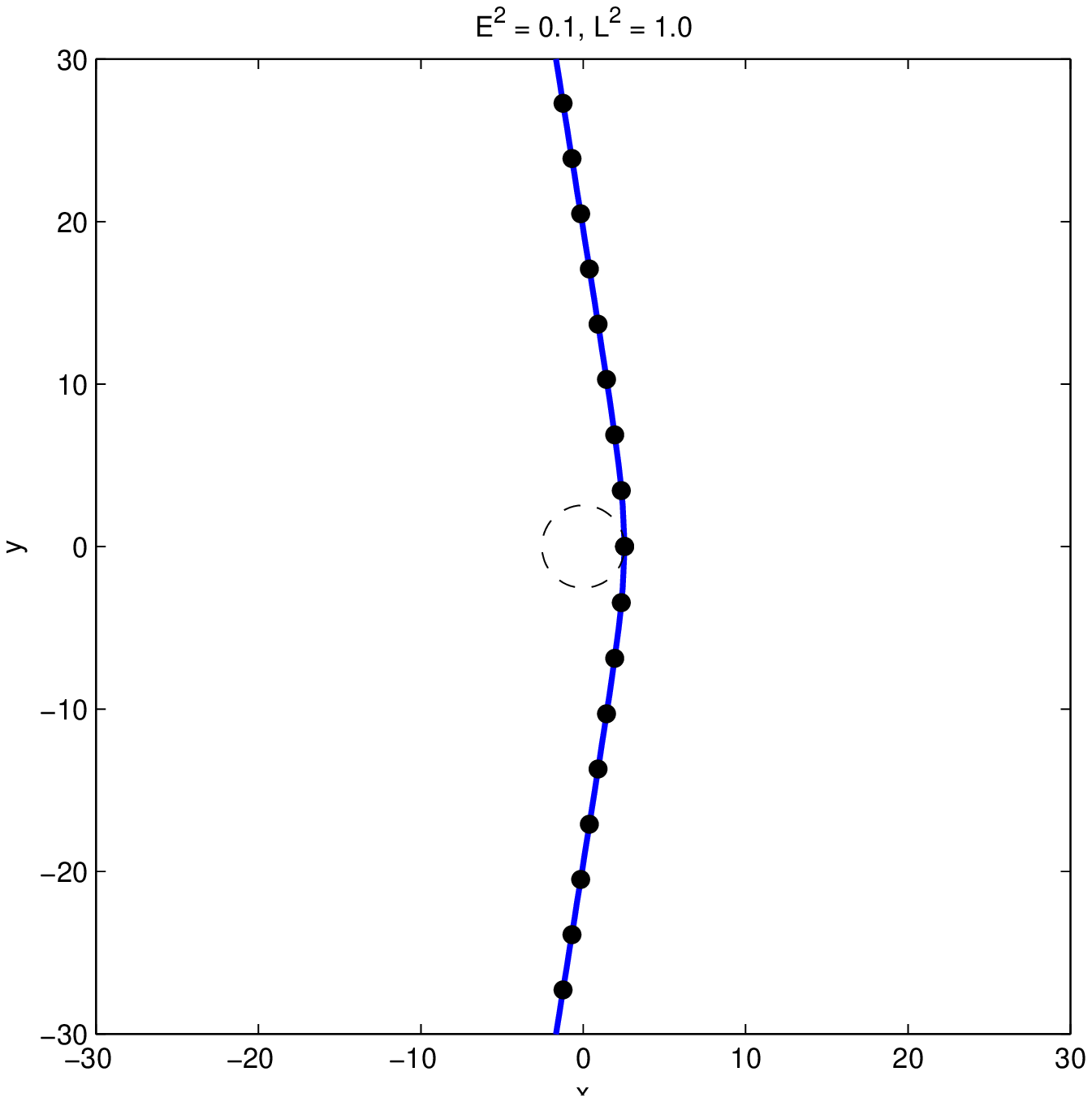}}
\subfigure[$\omega=0.8820$, $\kappa=0.1$]{\label{mo2}
\includegraphics[width=5.7cm]{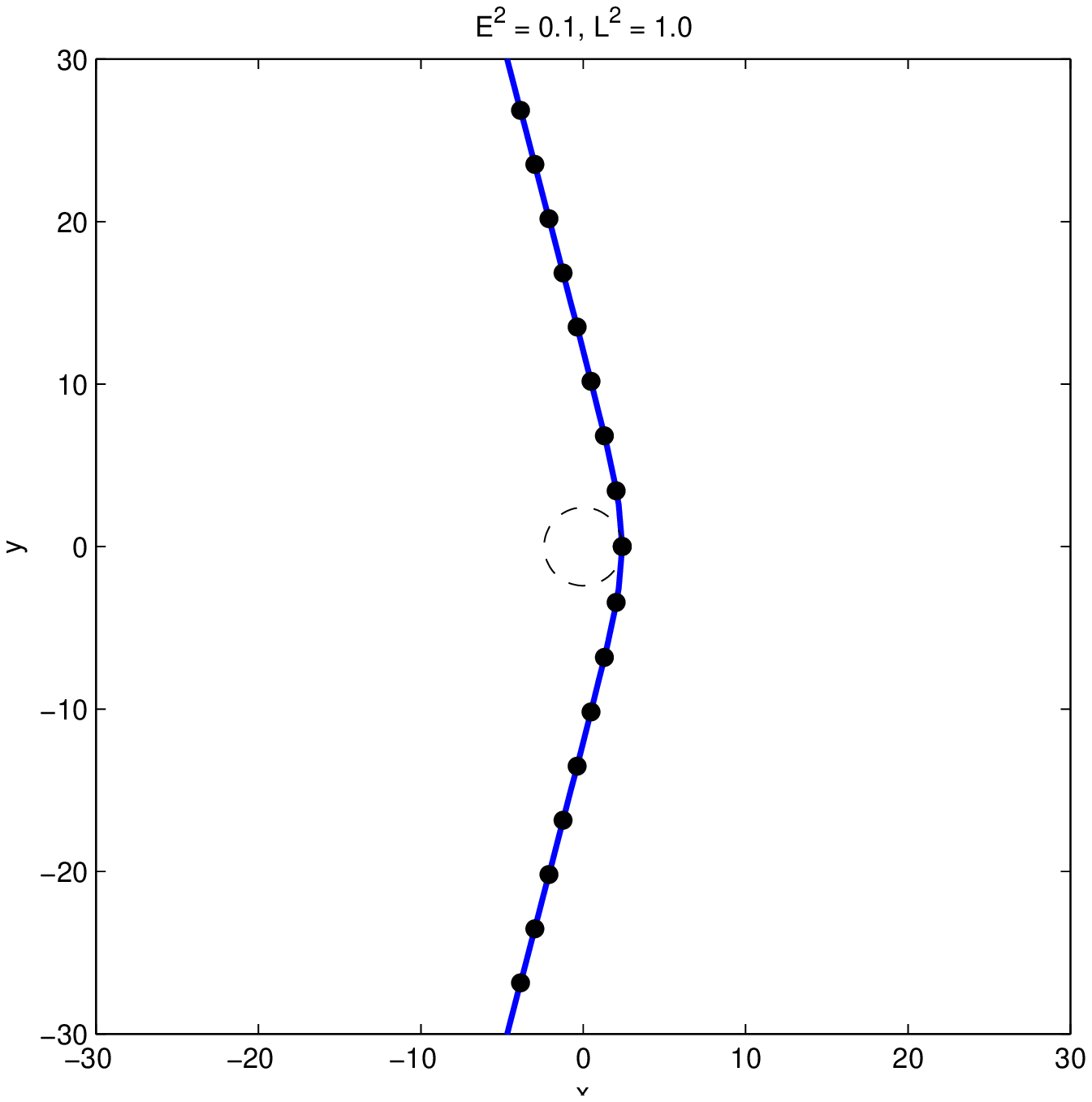}}\\
\subfigure[$\omega=0.6800$ (1), $\kappa=0.1$ ]{\label{mo3}
\includegraphics[width=5.7cm]{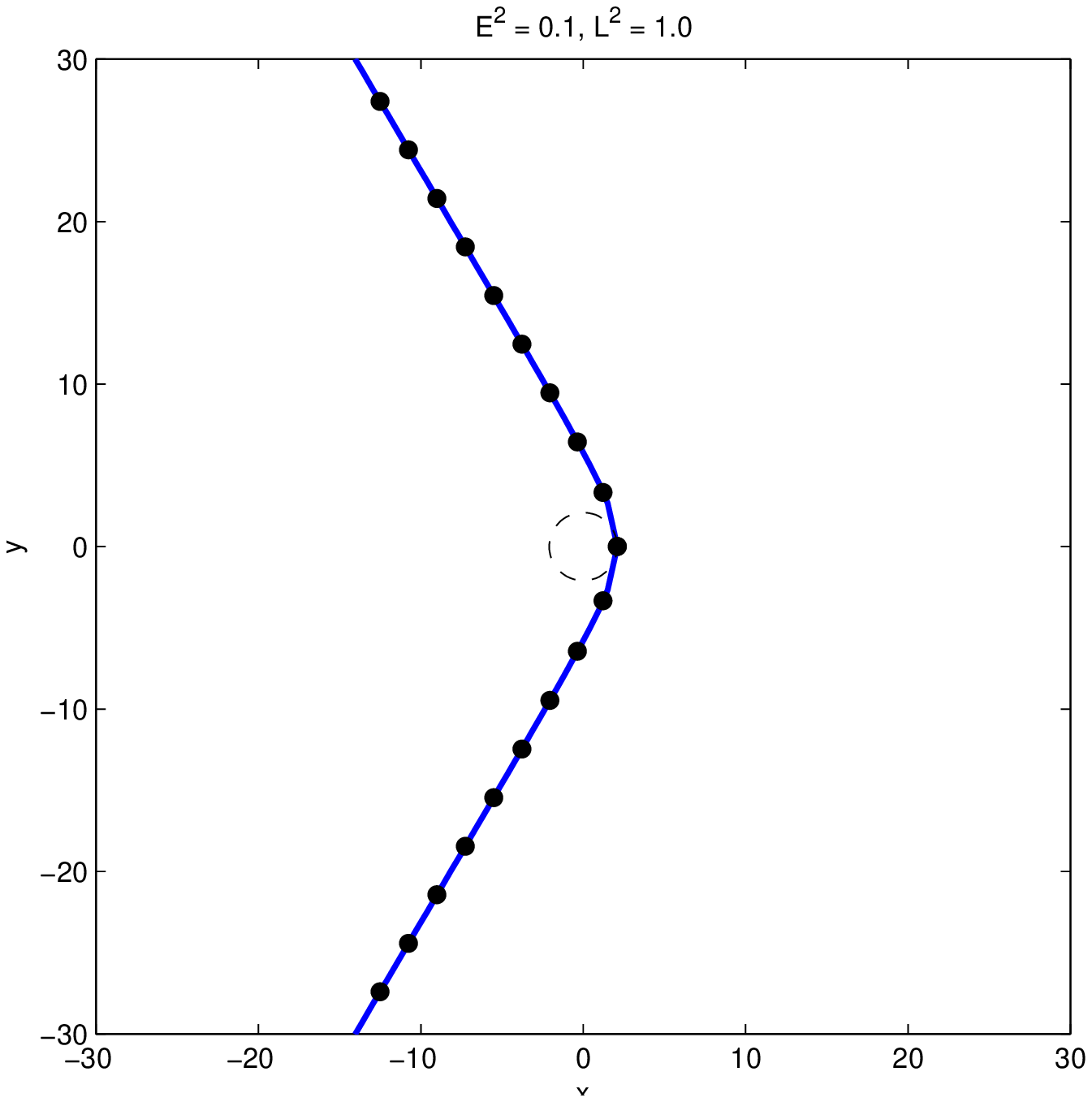}}
\subfigure[$\omega=0.6525$, $\kappa=0.1$]{\label{mo4}
\includegraphics[width=5.7cm]{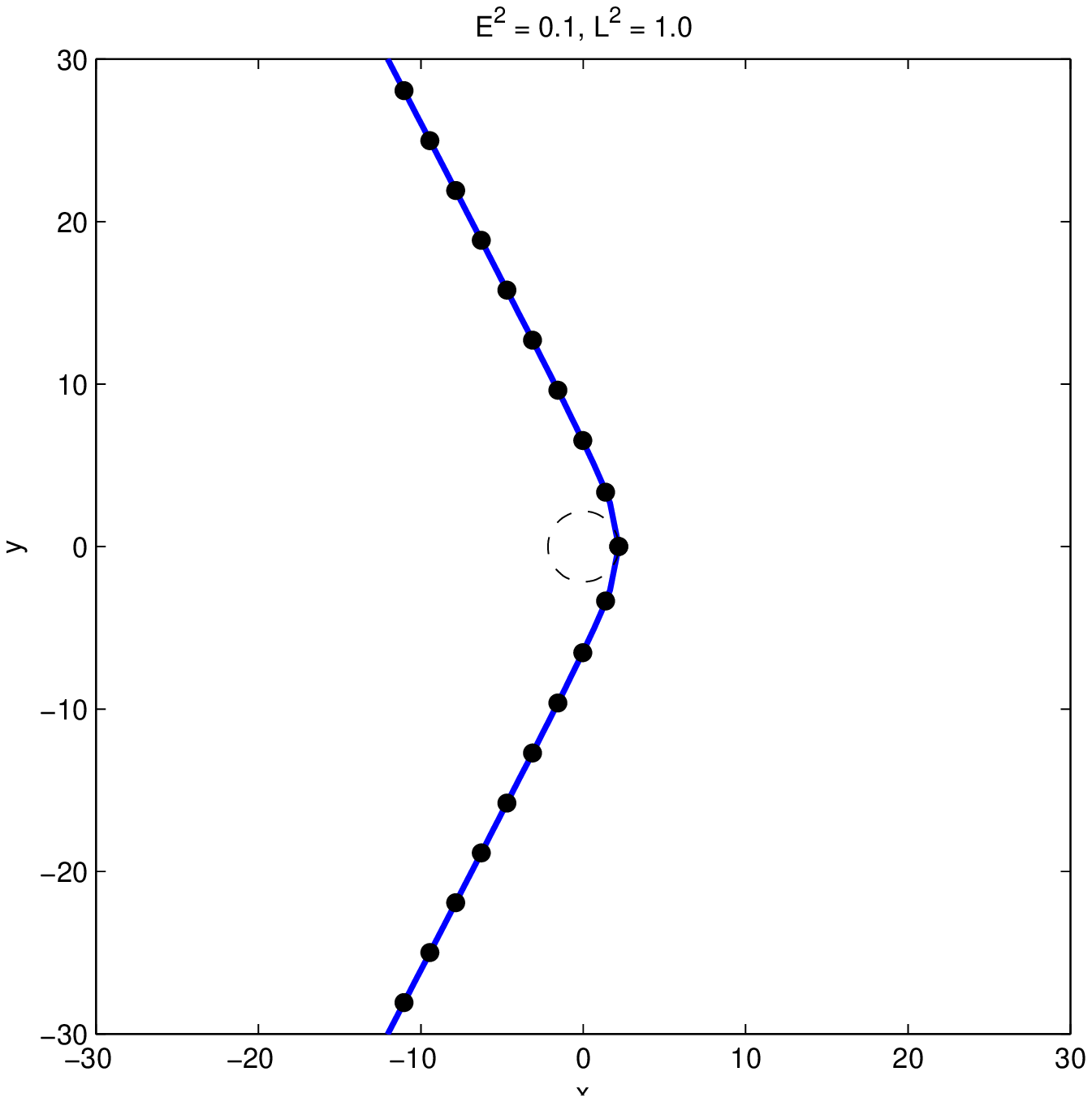}}\\
\subfigure[$\omega=0.6800$ (2), $\kappa=0.1$]{\label{mo5}
\includegraphics[width=5.7cm]{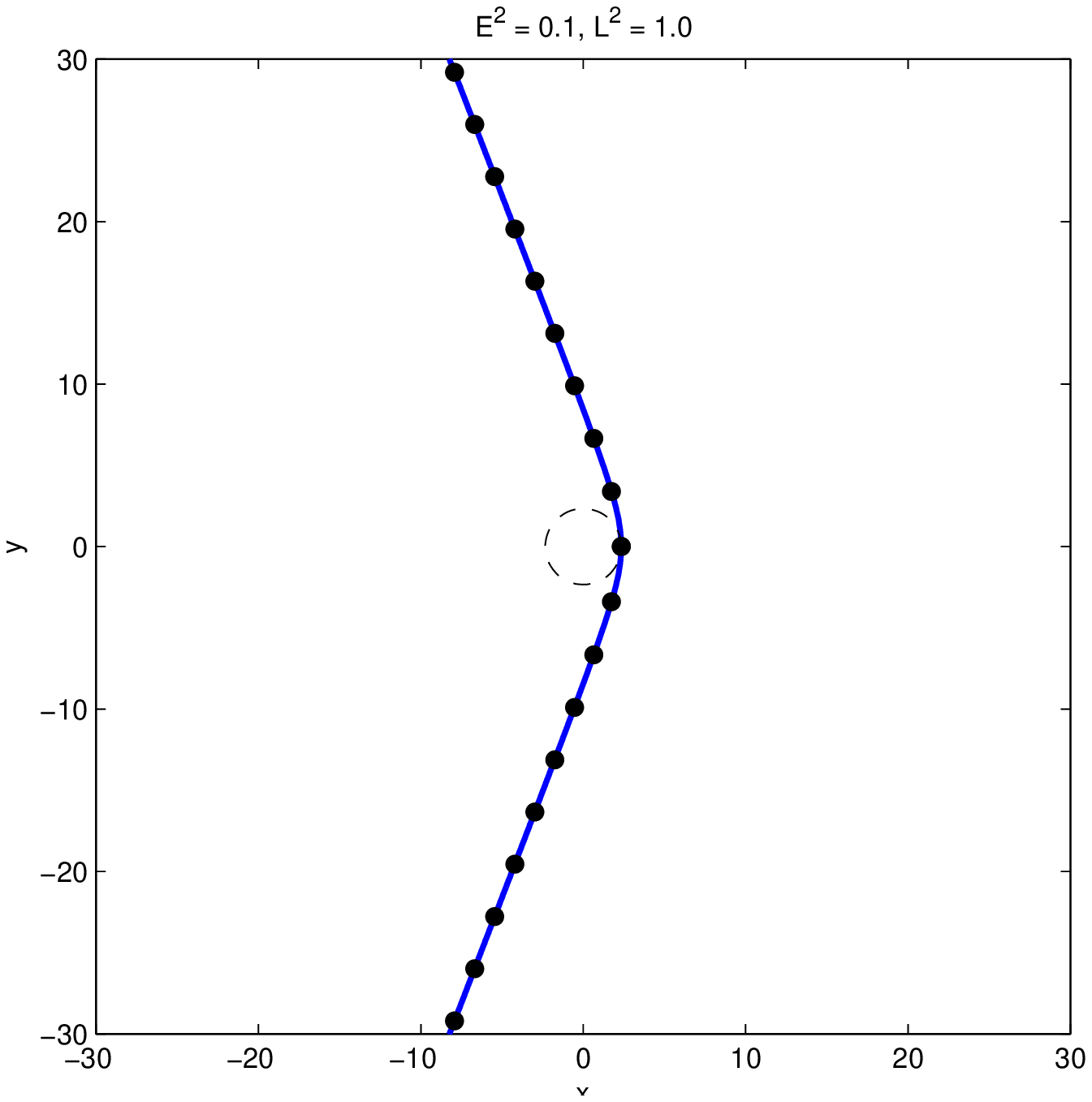}}
\subfigure[$\omega=0.8600$, $\kappa=1.0$ ]{\label{mo6}
\includegraphics[width=5.7cm]{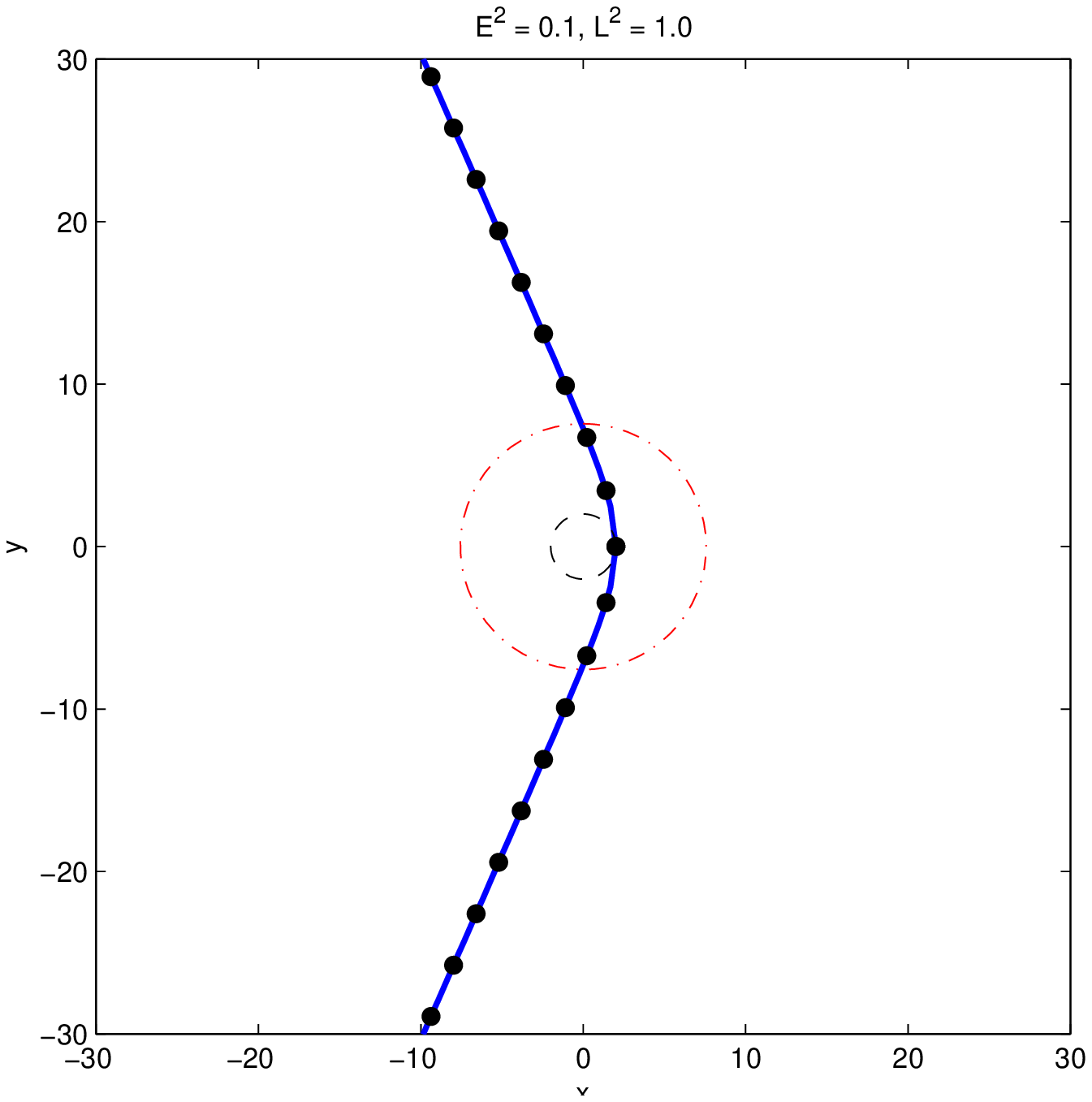}}\\
\subfigure[$\omega=0.76200$, $\kappa=1.0$ ]{\label{mo7}
\includegraphics[width=5.7cm]{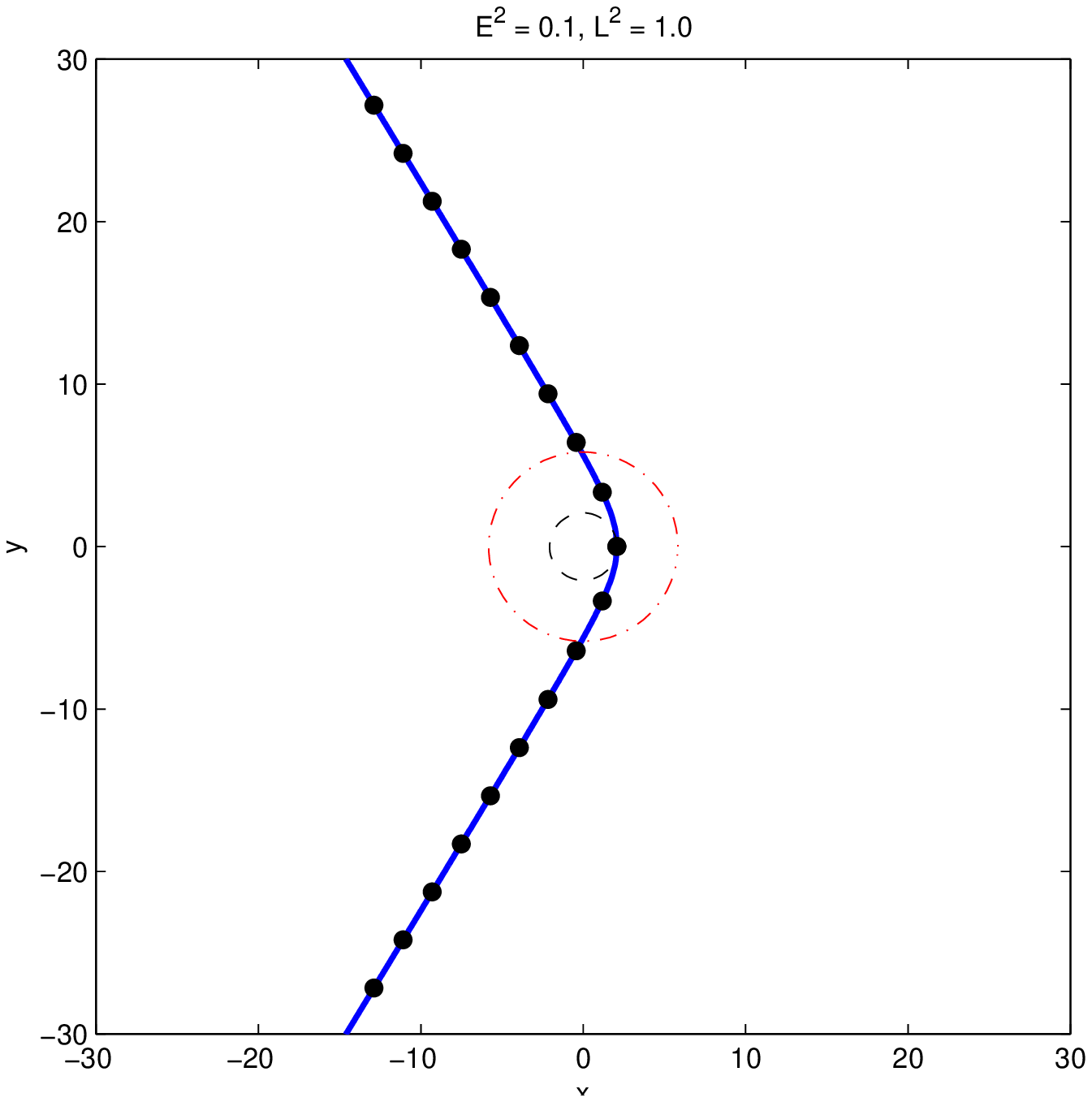}}
\subfigure[$\omega=0.8500$, $\kappa=1.0$ ]{\label{mo8}
\includegraphics[width=5.7cm]{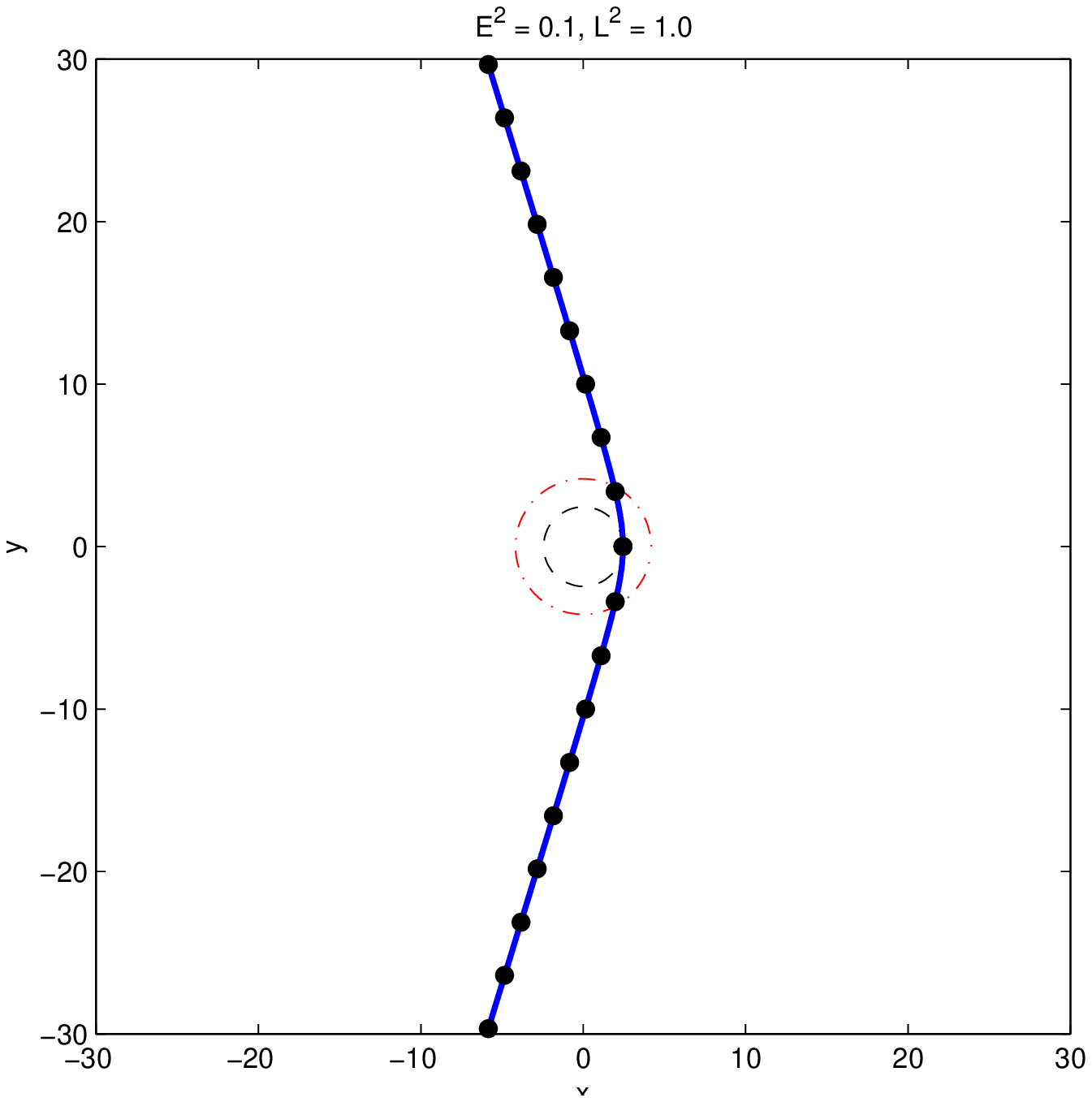}}
\caption{\label{massless_orbits} We show escape orbits of massless test particle with $E^2=0.1$ and $L^2=1.0$ 
in the space-time of a boson star
with $\kappa=0.1$ and $\kappa=1.0$, respectively for the boson stars given in Table \ref{omega_values}. Also indicated
is the radius of the event horizon of a Schwarzschild black hole with the same mass as that of the boson star 
(red circle) if this radius is larger than the minimal radius of the escape orbit, which is the case only for 
$\kappa=1.0$. The evolution of the proper time along the geodesic is shown by points.}    
\end{figure}

In Fig.\ref{massless_orbits} we show the escape orbits of massless test particles with $E^2=0.1$, $L^2=1.0$
in the space-time of the boson stars given in Table \ref{omega_values}. 
Depending on the choice of $\omega$ and $\kappa$ we find that the particles get deflected into a direction between
roughly $\pi/4$ and $\pi/2$ and that for $\kappa$ large enough the minimal radius of the escape orbit
lies inside the Schwarzschild radius of the corresponding black hole with the same mass. 
We do not plot terminating escape orbits of massless test particles here since they would just correspond to a line from
infinity to $r=0$.

\subsection{Observations of boson stars}
Next to the possibility that boson stars could act as a toy model for neutron stars, boson stars have been suggested 
as alternatives to supermassive black holes residing 
in the center of galaxies \cite{schunck_liddle}. As mentioned above, objects with a well-defined surface as 
alternatives to the
supermassive black hole at the center of our own Milky Way have been ruled out \cite{Broderick:2005xa}. 
We have, however
argued above that our boson stars are non-compact in the sense that there is no well-defined surface beyond which the
energy density and pressure, respectively, of the boson star strictly vanishes. 

\subsubsection{Boson star at the center of our galaxy}
The center of our Milky Way possesses a very compact astronomical radio source named {\it Sagittarius A$^*$}. 
It is believed to be a supermassive black hole as observations of the motion of stars around the galactic center
suggest that the mass of the central object is on the order of $\hat{M}\approx 10^{37} kg$ which corresponds
to roughly $4\cdot 10^6$ solar masses \cite{Ghez:2008ms,Gillessen:2008qv}, which is contained within a sphere of radius
roughly $2\cdot 10^{10} m$. Comparing with Table \ref{table_particles} we observe that we can have boson stars
that can be as dense as the observed object. Furthermore, since the orbits of a number of stars 
have been determined, we can compare the observational results with ours. Unfortunately, the perihelion shift of
the orbits of stars around the galactic center has not yet been determined (further observations are necessary here),
but the eccentricity $e:=\sqrt{(r_{\rm max}^2 - r_{\rm min}^2)/r_{\rm max}^2}$ of the 
orbits has been calculated from the observations. In Table \ref{rmin_rmax} we give some values of $e$ for orbits
in boson star space-times. Note that these are the nearly maximally available eccentricities, but that we can
also have orbits which are nearly circular (see discussion above on SCOs). Hence $e \geq 0$ up to $e\lesssim 1$ is possible.
Interestingly, some of the orbits of stars circling around the center of our own Milky Way have quite large
eccentricities \cite{Ghez:2003qj,Eisenhauer:2005cv, Meyer:2012hn}, e.g. the orbit of the star called $S14$ could 
have eccentricity up to $e\approx 0.974\pm 0.016$ \cite{Ghez:2003qj}. Clearly, we are able to describe these large
eccentricities with our model. 

\begin{figure}[p!]
\subfigure[bound orbit, $\omega=0.8600$]{\label{kappa1_bound1}
\includegraphics[width=6cm]{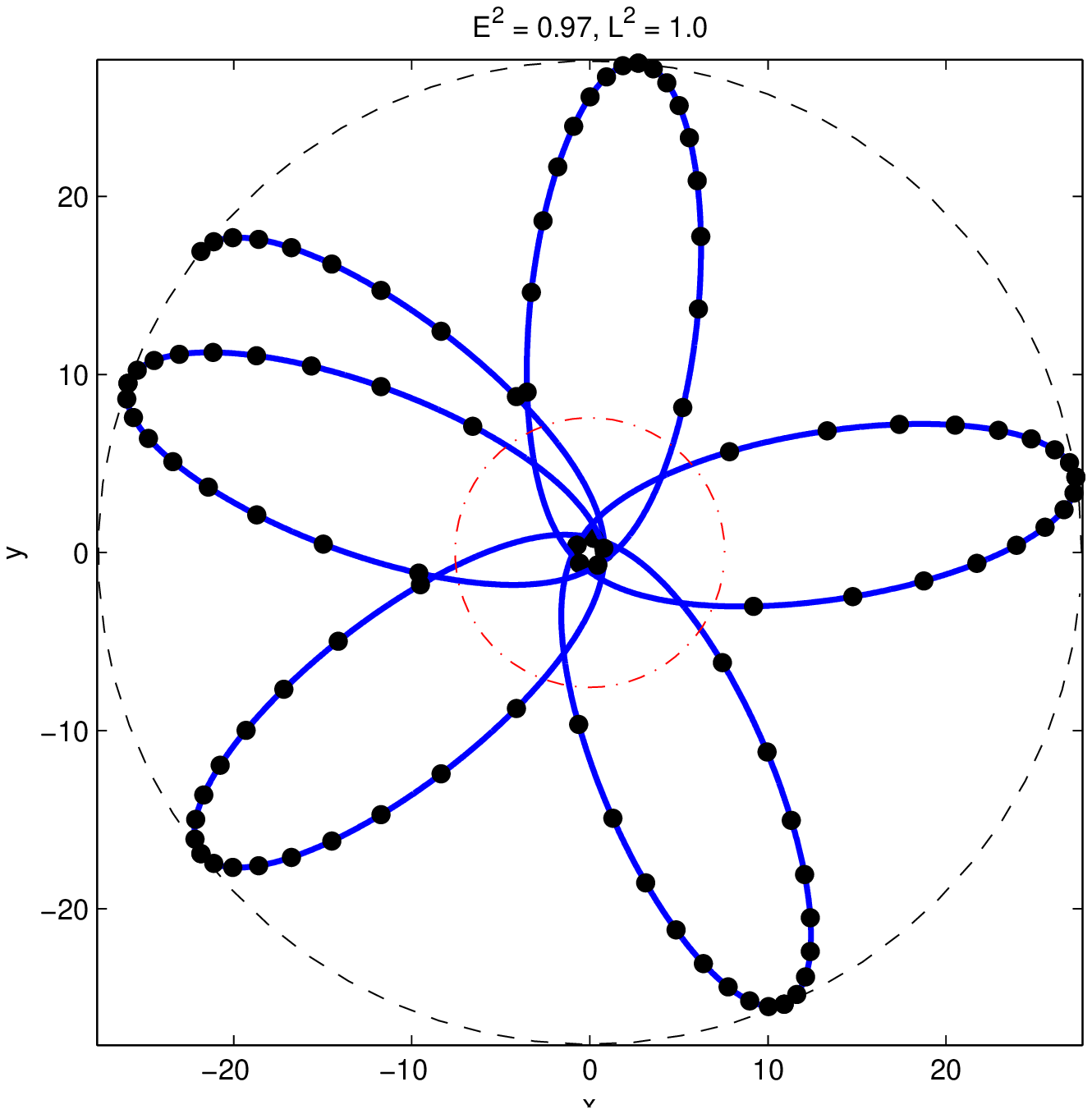}}
\subfigure[escape orbit, $\omega=0.8600$]{\label{kappa1_escape1}
\includegraphics[width=6cm]{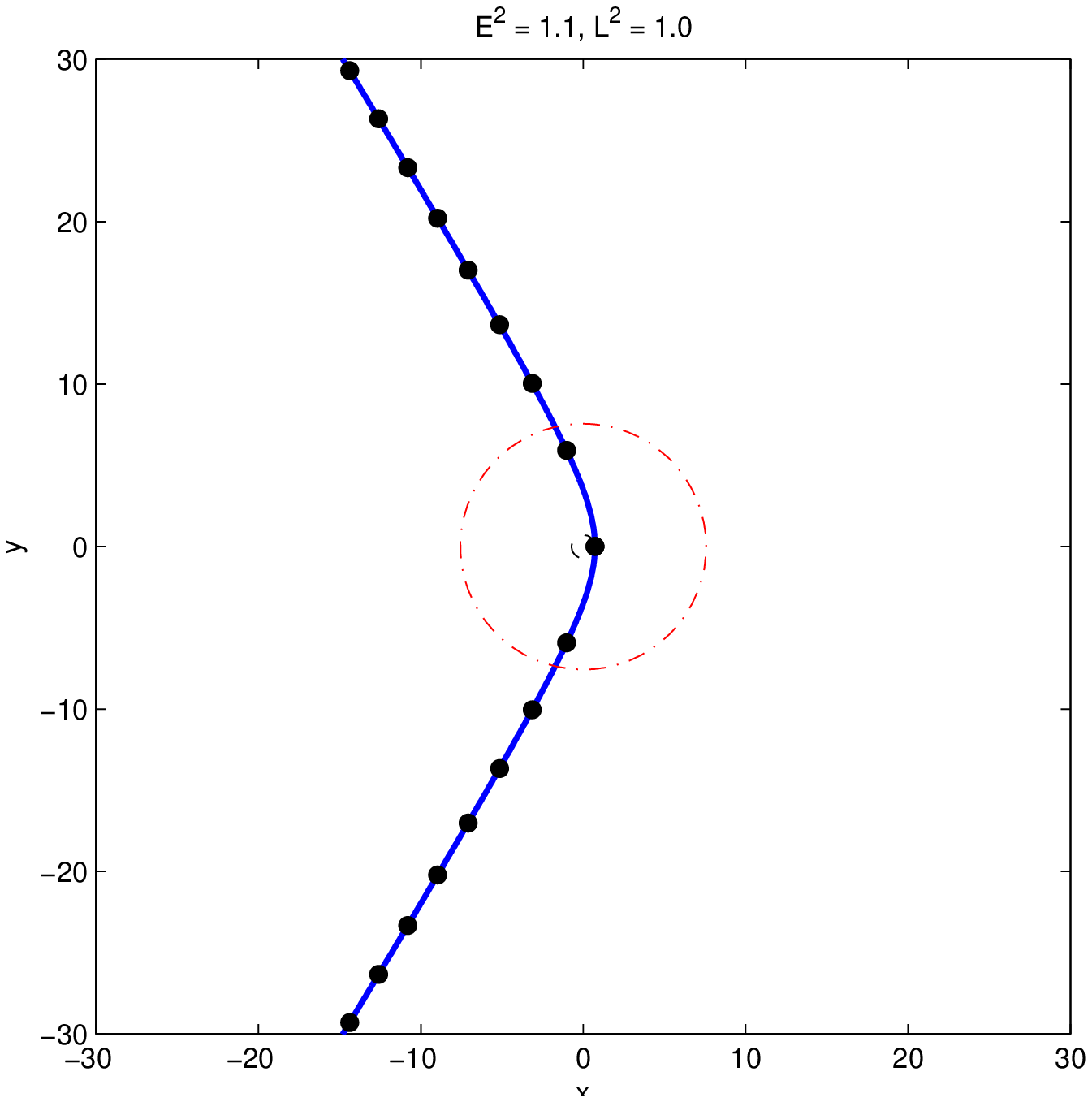}}\\
\subfigure[bound orbit, $\omega=0.8500$  ]{\label{kappa1_bound2}
\includegraphics[width=6cm]{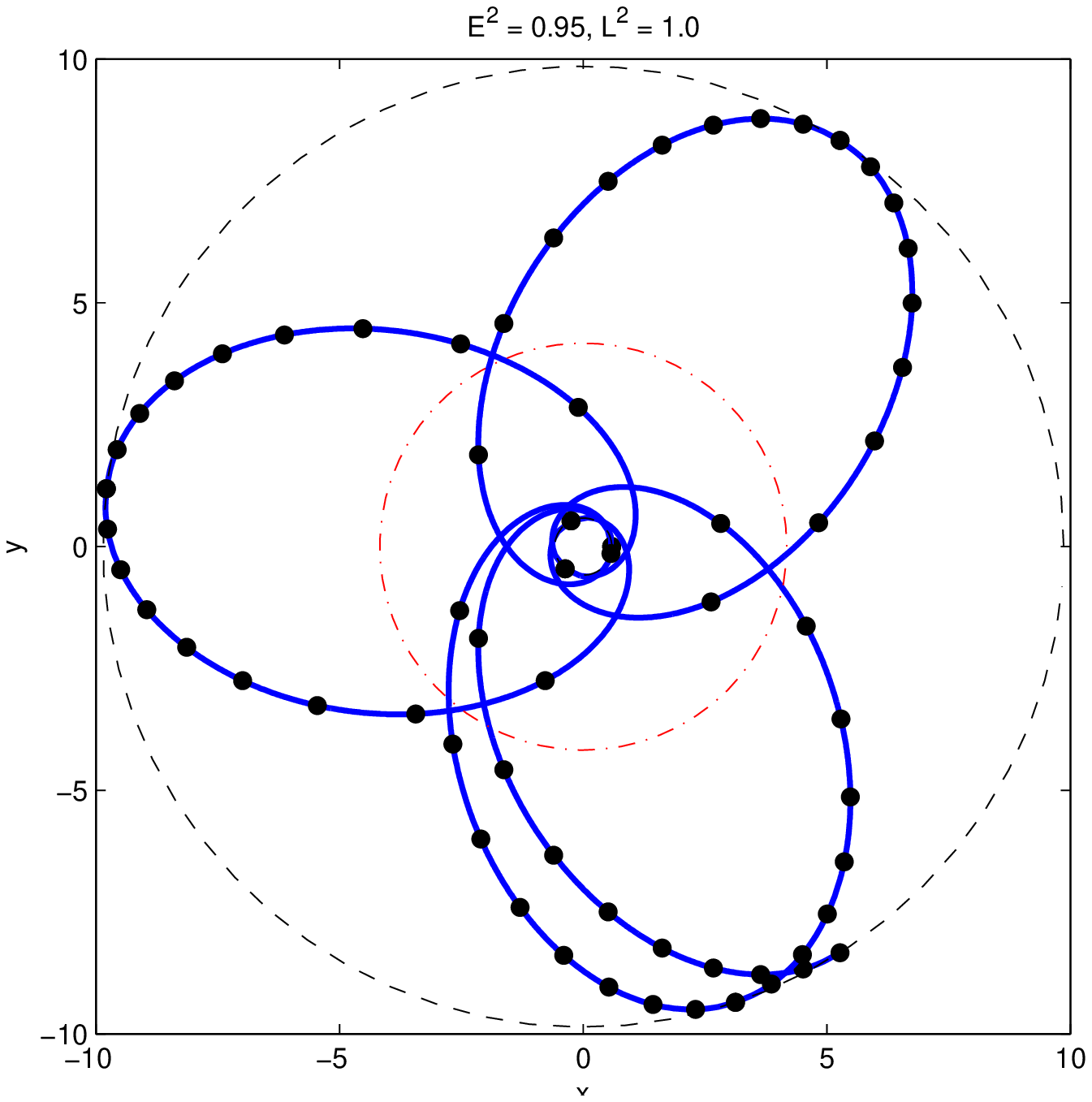}}
\subfigure[escape orbit, $\omega=0.8500$ ]{\label{kappa1_escape2}
\includegraphics[width=6cm]{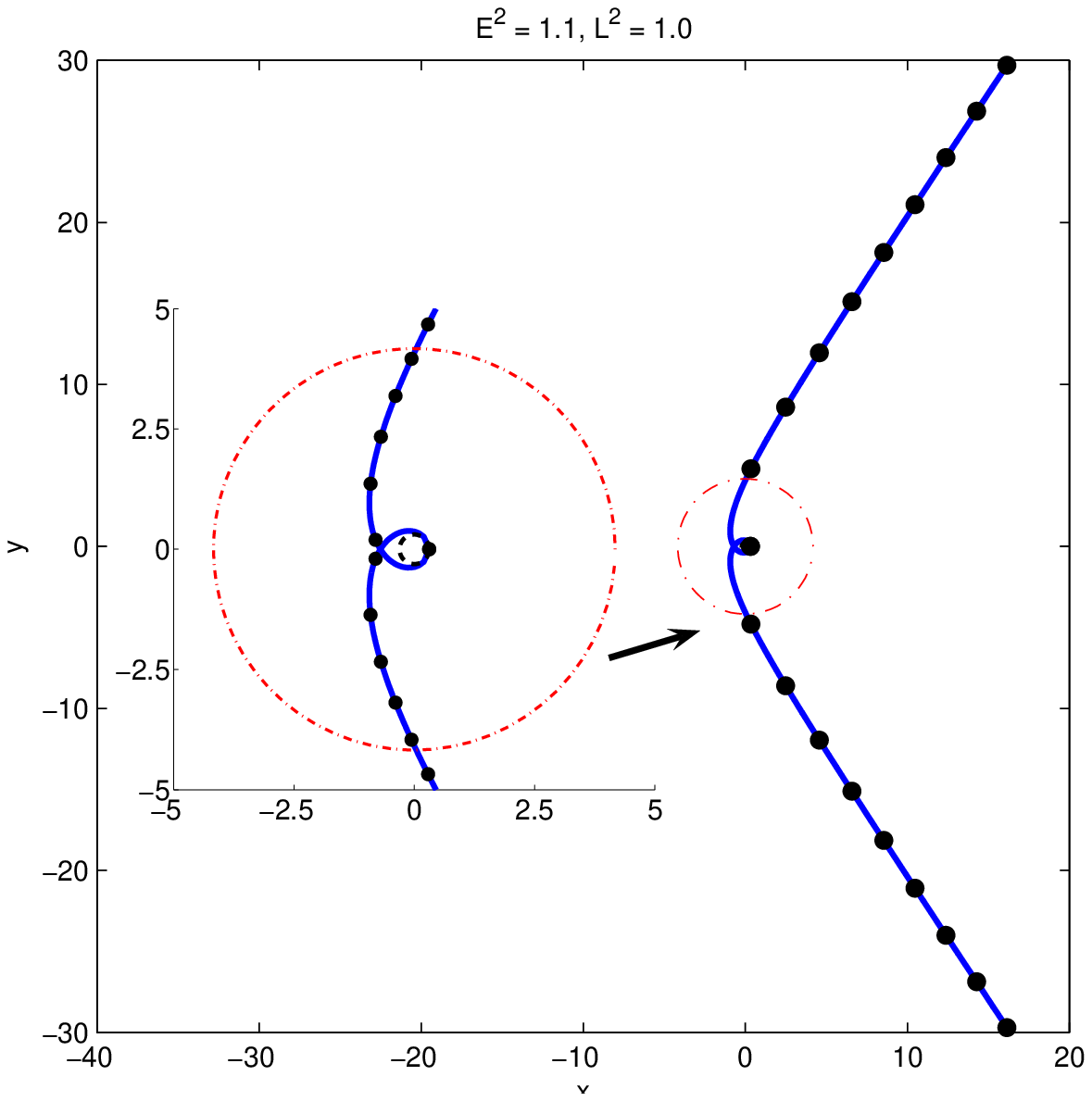}}\\
\subfigure[bound orbit, $\omega=0.7620$]{\label{kappa1_bound2}
\includegraphics[width=6cm]{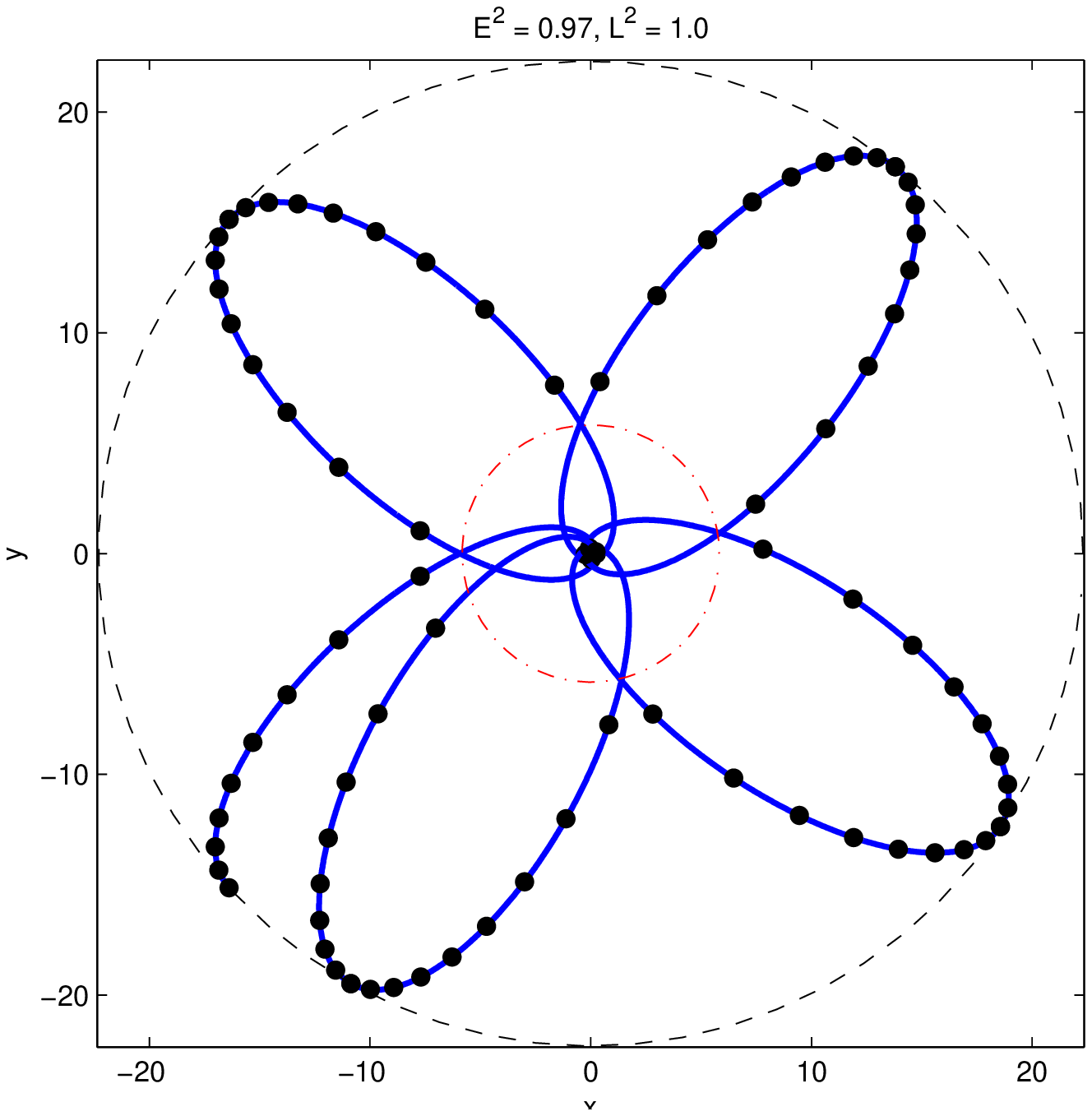}}
\subfigure[escape orbit, $\omega=0.7620$  ]{\label{kappa1_escape2}
\includegraphics[width=6cm]{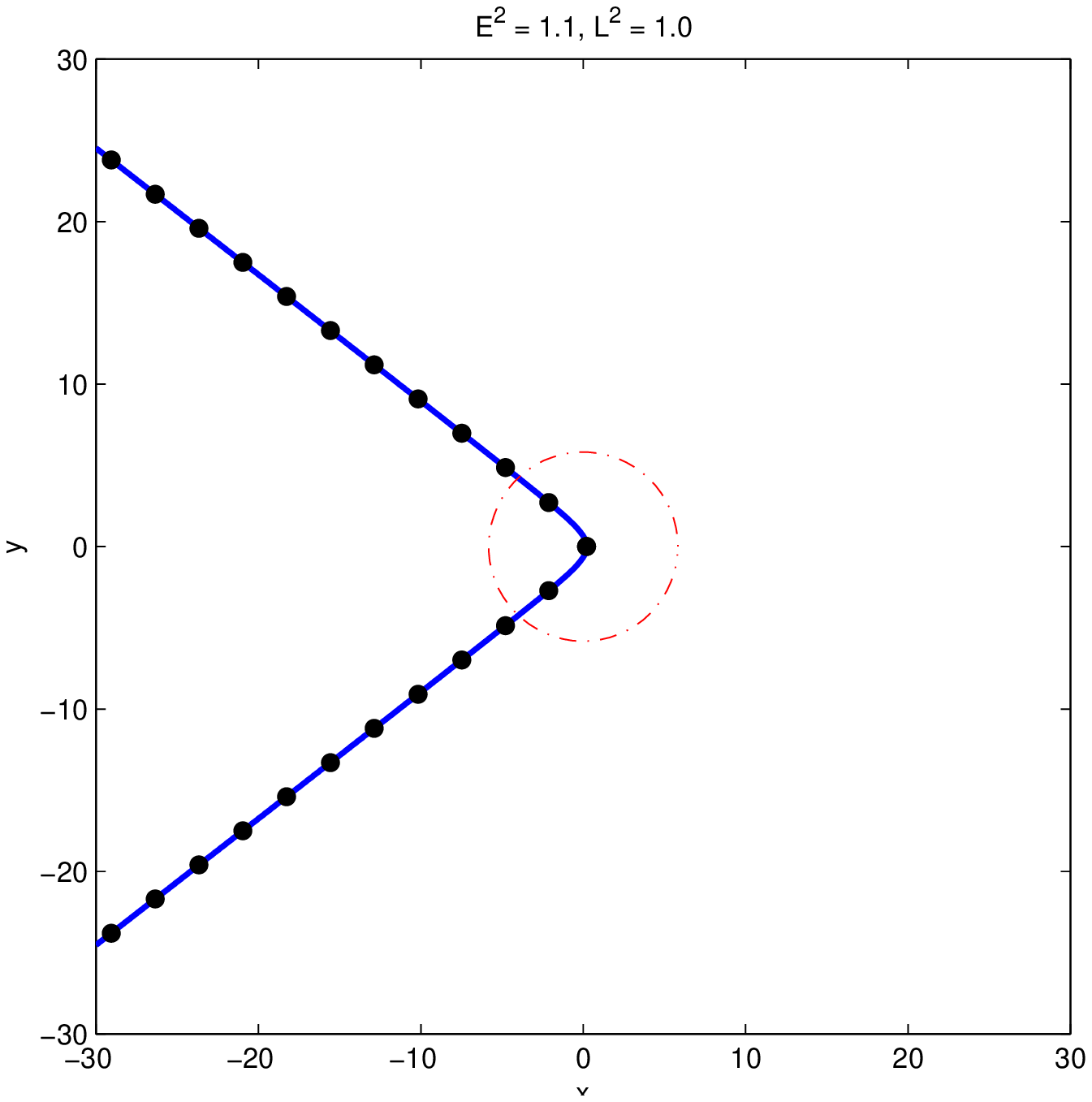}}\\
\caption{\label{kappa1} We show bound and escape orbits of massive test particle in the space-time of a boson star
with $\kappa=1.0$ for different values of $\omega$. Also indicated is the Schwarzschild radius $r_h=M/2$ (red circle), 
i.e. the radius of the event horizon of a black hole with the mass  $M$ equal to that of the respective boson star (see also Table \ref{omega_values}). The evolution of the proper time along the geodesic is shown by points. A test particle covers a larger distance in the same period of proper time, i.e. moves faster, in the vicinity of its perihelion.}    
\end{figure}

Another interesting observation has been done related to the emission of short wavelength radio
waves from {\it Sagittarius A$^*$}. These are emitted due to accretion
of surrounding matter onto the supermassive object in the center of our Milky Way.  In \cite{Doeleman:2008qh}
radio waves with wavelength of $1.3$ mm have been observed in order to understand the
region close to the supermassive object in which gravitational effects should be very strong.
Observing {\it Sagittarius A$^*$} at wavelength $1.3$ mm, the authors of \cite{Doeleman:2008qh} concluded  
that the size of this object is smaller than the expected apparent event horizon size of the assumed black hole. 
It was suggested that this might be due to the fact that the biggest fraction of emission of radio waves
is not due to the black hole at the center, but rather related to the accretion in the vicinity of the supermassive
object. 

If we would, however, give up the idea of a black hole in the center of our galaxy and use a non-compact boson
star instead particles could be emitted from radii smaller than the apparent event horizon of the black hole
since the space-time is globally regular. This is demonstrated in Fig.\ref{kappa1}, where we show
bound and escape orbits of massive test particles in the space-time of a boson star with $\kappa=1.0$ and
different values of $\omega$. For comparison we also show a circle with radius equal to 
the Schwarzschild radius of a black hole
with the same mass, which in our coordinates is $r_h=M/2$. It is apparent that the minimal radius of these
orbits (both bound as well as escape) is smaller than the event horizon, while the maximal radius of
the bound orbits is larger. If we would hence have stricter bounds on the mass (and hence the corresponding
Schwarzschild radius) of the supermassive
object in the center of our galaxy and would have clear signals from particles coming
from ``inside'' the event horizon a possible explanation might be a non-compact boson star.
Future observations at wavelength even smaller than $1.3$ mm should be able
to show what the space-time of the supermassive object at the center of our galaxy looks like. 

\subsubsection{Perihelion shift}
The motion of a number of stars orbiting the center of our Milky Way have been determined 
\cite{Ghez:2003qj,Eisenhauer:2005cv, Meyer:2012hn}. Since the orbital periods $T_o$ of these stars are quite large
(between $T_o=11.5\pm 0.3$ years for S0-102 \cite{Meyer:2012hn} and $T_o=94.1\pm 9.0$ years for S1 \cite{Eisenhauer:2005cv})
it will need longer observations to determine the perihelion shift of these objects. But this is certainly one of 
the interesting observables. Hence, we discuss the possible perihelion shifts of the orbits of massive
test particles in the space-time of non-compact boson stars here. These might be later used to compare with observations.

The perihelion shift of a bound orbit of a massive test particle is given by
\begin{equation}
\label{pshift}
 (\Delta\varphi)_{\delta=1} = 2
\int\limits_{r_{\rm min}}^{r_{\rm max}} \frac{dr}{\sqrt{E^2 - V_{\rm eff}(r)}} \frac{L f}{r^2 \sqrt{l}} -2\pi \ .
\end{equation}
with $V_{\rm eff}$ given by (\ref{effective}), where we set $\delta=1$.

   \begin{table}  \centering
			
    \begin{tabular}{|c|c|c|c|c|c|c|}
\hline
    $\omega$ & $\kappa$  &   $E^2$     &    $L^2$    &      $(\Delta\varphi)_{\delta=1}$   \\ \hline\hline
   		    0.9260 & 0.1       &   0.97      &    1.0      &      $-0.4830\pi$  \\ \hline
	   0.8820  & 1.0  &   0.95      &    1.5      &      $-0.1270\pi$  \\ \hline	
 	0.6800 (1)  & 0.1    &   0.95      &    0.5      &      $-0.1658\pi$  \\ \hline
0.6525 & 0.1       &   0.95      &    1.5      &      $1.3829\pi$  \\ \hline
 	0.6800 (2) & 0.1       &   0.95      &    0.5      &  $0.5408 \pi$  \\  \hline
	0.8600 & 1.0         &   0.97      &    1.0      &      $-0.4196\pi$  \\ \hline
			 0.7620 & 1.0      &   0.97      &    1.0      &    $0.4749\pi$  \\  \hline             
				0.8500 & 1.0          &   0.95      &    1.0      &      $1.3590\pi$  \\ \hline          

	\end{tabular}
 \caption{The values of the perihelion shift $(\Delta \varphi)_{\delta=1}$ for the boson star space-times given in Table \ref{omega_values}
and for the orbits with nearly maximal eccentricity (see also Table \ref{rmin_rmax}). }
\label{perihelion}
	\end{table}

The values for orbits with nearly maximal eccentricity are given in Table \ref{perihelion}. We observe that
the perihelion shift in all cases is quite large and can even become negative. This latter fact would be something
that is not possible in a Schwarzschild black hole space-time and would allow us to distinguish black holes
from boson stars. 

\subsubsection{Light deflection}
Since the gravitational field close to a supermassive object is very strong it is important to understand
light deflection in order to be able to interpret observational data correctly. Due to the strong bending of light rays
the apparent event horizon of the proposed black hole would be twice as big as the actual event horizon.
All of these things have to be taken into account when analyzing observational data.

The light deflection in the space-time of a boson star is given by
\begin{equation}
\label{pshift}
 (\Delta\varphi)_{\delta=0} = 2
\int\limits_{r_{\rm min}}^{\infty} \frac{dr}{\sqrt{E^2 - V_{\rm eff}(r)}} \frac{L f}{r^2 \sqrt{l}} - \pi \ .
\end{equation}
with $V_{\rm eff}$ given by (\ref{effective}), where we set $\delta=0$.

   \begin{table}  \centering
			
    \begin{tabular}{|c|c|c|c|c|c|c|}
\hline
    $\omega$ & $\kappa$ &   $E^2$     &    $L^2$    &      $(\Delta\varphi)_{\delta=0}$   \\ \hline\hline 
  0.9260 & 0.1        &   0.1      &    1.0      &      $0.0820\pi$  \\ \hline				
		   0.8820 & 0.1       &   0.1      &    1.0      &      $0.1444\pi$  \\ \hline	
  	0.6800 (1)   & 0.1       &   0.1      &    1.0      &      $0.3188\pi$  \\ \hline	
   0.6525 & 0.1        &   0.1      &    1.0      &      $0.2846\pi$  \\ \hline
0.6800 (2) & 0.1       &   0.1      &    1.0      &      $0.2147\pi$ \\ \hline
	0.8600 & 1.0     &   0.1      &    1.0      &      $0.2449\pi$  \\ \hline
			 0.7620 & 1.0           &   0.1      &    1.0      &      $0.3281\pi$  \\ \hline
				0.8500 & 1.0 &   0.1      &    1.0      &      $0.1697\pi$\\  \hline
			
	\end{tabular}
 \caption{The values of the light deflection $(\Delta \varphi)_{\delta=0}$ for the boson star space-times given 
in Table \ref{omega_values} and $E^2=0.1$ and $L^2=1.0$.  }
\label{light_deflection}
	\end{table}

In Table \ref{light_deflection} we give the light 
deflection in the boson star space-times indicated in Table \ref{omega_values} for massless test particles
with $E^2=0.1$ and $L^2=1.0$. 
We observe that the light deflection increases when moving along the branch for $\kappa=0.1$ up to
roughly the minimal possible value of $\omega\approx 0.6525$. On the second branch, the light deflection decreases
again. Similar to what was  stated above for massive test particles, there seems to be no direct correlation between the mass
of the boson star and the value of light deflection and hence the presence of the scalar field seems to have a
non-trivial effect. For $\kappa=1.0$ we find that the light deflection decreases when moving along the branches.
This again indicates that a semi-analytical treatment of the space-time of boson stars does not lead
to the proper results and that numerical techniques are vital to make predictions about observables in such kind of
space-times. 

\section{Conclusions and Outlook}
In this paper, we have studied the geodesic motion of test particles in boson star space-times and have discussed 
possibilities to 
differ them from the geodesics around a Schwarzschild black hole. Our work has many interesting applications.
As a first example let us mention the accretion of matter onto massive compact objects such as neutron stars.
Neutron stars are very dense objects that are typically modeled using an appropriate equation of state.
Boson stars could act as a toy model for these objects. In the following, one could use our techniques
to model accretion flow onto a boson star in a simple ballistic approach. This was done before in black hole
space-times where analytic solutions to the geodesic equation exist \cite{Tejeda:2011pr,Tejeda:2012kb}.
This technique could be applied as well using the numerical data of a boson star space-time.
The accretion flow would also be interesting from the point of view of interpreting the boson star as an alternative
to a supermassive black hole residing in the center of our Milky Way. The accretion of matter leads to
the emission of radio waves and these can be used to make predictions about the properties of the supermassive
object at the center of our galaxy. Gravitational wave signals would be another possibility. These have already
been discussed in \cite{Kesden:2004qx}. 

It would also be interesting to see how test particles move in radially excited boson star space-times 
as well as in the space-time of a rotating boson star. For once this is interesting since it is believed that
the flat space-time counterparts of boson stars, so-called $Q$-balls could originally form in an excited 
state \cite{kusenko}. Furthermore, due to the interaction with surrounding matter, e.g. accretion, boson stars
might be excited. 
If boson stars should act as toy model for neutron stars it would be surely also of interest to study
the rotating counterparts since basically all observed neutron stars seem to rotate, in fact quite
quickly as the example of a radio pulsar (assumed to be a neutron star) spinning at 716 Hz shows 
\cite{Hessels:2006ze}. It would then be interesting to study the motion of test particles in these space-times.

Finally, in order to model accretion flow onto dense astrophysical objects more realistically one should
consider charged test particles (that would make up the plasma). Hence the equation of motion for the particle
should be modified to include the charge of the particle. Studies of charged test particles in some black hole
space-times have been done previously \cite{charged_test}. In order to have interactions of the test particles
with the boson star the latter should be charged as well. This can be done by promoting the global U(1)
symmetry of our model to a local U(1) and has been considered for a number of boson stars \cite{boson_charged}. 
The corresponding equation of motion for the test particle would then
have to be solved numerically. \\
\\
\\

{\bf Acknowledgment} We gratefully acknowledge the Deutsche Forschungsgemeinschaft (DFG) for financial support
within the framework of the DFG Research Training group 1620 {\it Models of gravity}. We also acknowledge 
discussions with Parinya Sirimachan at the initial stages of this paper.


\begin{thebibliography}{99}
\bibitem{footnote1} In the following, we will denote
dimensionful quantities with a hat.
\end{thebibliography}
\end{document}